\documentclass[a4paper,fleqn,usenatbib]{mnras}

\usepackage{newtxtext,newtxmath}

\usepackage[T1]{fontenc}
\usepackage{ae,aecompl}

\usepackage{graphicx,xcolor}	
\usepackage{amsmath}	
\usepackage{amssymb}	

\newcommand{\cormaa}[1]{{\color{violet}  [MAA: {\bf #1}]}}

\newcommand*{\rplc}[2]{{\color{red} \st{#1}}{\color{blue} #2}}

\newcommand{\dsfrac}[2]{\displaystyle{\frac{#1}{#2}}}

\def\Ljth{L_{\rm  j}^{\rm thr}}

\newif\iflinenumbers

\iflinenumbers
  \usepackage{lineno}
  \modulolinenumbers[5]
\fi

\usepackage{soul}
\soulregister\cite7
\soulregister\citep7
\soulregister\citealt7
\soulregister\ref7
\soulregister\figref7
\soulregister\tabref7
\soulregister\eqref7
\soulregister\st7
\soulregister\rplc7
\soulregister\corr7
\soulregister\cormaa7

\title[Luminosity threshold of GRB jets in massive stars]{On the
  existence of a luminosity threshold of GRB jets in massive stars}

\author[M. A. Aloy, C. Cuesta-Mart\'inez, \& M. Obergaulinger]{
M. A. Aloy\thanks{E-mail: miguel.a.aloy@uv.es (MAA)}
C. Cuesta-Mart\'inez,\thanks{E-mail: carlos.cuesta@uv.es (CC-M)}
and M. Obergaulinger\thanks{E-mail: martin.obergaulinger@uv.es (MO)}
\\
Departamento de Astronom\'ia y Astrof\'isica, Universidad de Valencia, \\
Edificio de Investigaci\'on Jeroni Munyoz, C/ Dr. Moliner, 50, E-46100 Burjassot (Valencia), Spain
}

\date{Accepted XXX. Received YYY; in original form ZZZ}

\pubyear{2018}

\begin{document}
\label{firstpage}
\pagerange{\pageref{firstpage}--\pageref{lastpage}}
\maketitle

\begin{abstract}
  Motivated by the many associations of $\gamma$-ray bursts (GRBs)
  with energetic supernova (SN) explosions, we study the propagation
  of relativistic jets within the progenitor star in which a SN shock
  wave may be launched briefly before the jets start to propagate.
  Based on analytic considerations and verified with an extensive set
  of 2D axisymmetric relativistic hydrodynamic simulations, we have
  estimated a threshold intrinsic jet luminosity, $\Ljth$, for
  successfully launching a jet. This threshold depends on the
  structure of the progenitor and, thus, it is sensible to its mass
  and to its metallicity. For a prototype host of cosmological long
  GRBs, a low-metallicity star of 35 $M_{\odot}$, it is $\Ljth \simeq
  1.35\times 10^{49}$\,erg s$^{-1}$. The observed equivalent isotropic
  $\gamma$-ray luminosity, $L_{\rm \gamma,iso,BO} \simeq 4
  \epsilon_\gamma L_{\rm j} \theta_{\rm BO}^{-2}$, crucially depends
  on the jet opening angle after breakout, $\theta_{\rm BO}$, and on
  the efficiency for converting the intrinsic jet luminosity into
  $\gamma$-radiation, $\epsilon_\gamma$. Highly energetic jets can
  produce low-luminosity events if either their opening angle after
  the breakout is large, which is found in our models, or if the
  conversion efficiency of kinetic and internal energy into radiation
  is low enough. Beyond this theoretical analysis, we show how the
  presence of a SN shock wave may reduce this luminosity threshold by
  means of numerical simulations.
  We foresee that the high-energy transients released by jets produced
  near the luminosity threshold will be more similar to llGRBs or XRFs
  than to GRBs.
\end{abstract}



\iflinenumbers
   \linenumbers
\fi

\section{Introduction}

The association of long gamma-ray bursts (GRBs) with supernovae (SNe;
in particular with broad-lined Type Ic SNe, SNe Ic-bl) has been
observationally confirmed at low redshifts (typically, $z\lesssim 0.3$
\citealt{Galama_1998, Patat_2001,Hjorth_2003};
\citealt{Stanek_etal_ApJ_2003__SN2003dh_GRB030329,Malesani_2004,Pian_2006,Bufano_2012};
\citealt{Hjorth_2013,Modjaz_2016,Cano_2017}).  The detection of these
two events together plays in favor of the collapsar model
\citep{Woosley_1993, MacFadyen_Woosley_ApJ_1999__Collapsar} and points
towards Wolf-Rayet (WR) stars with masses $> 10 M_\odot$ as the
likeliest candidates to host the central engines of long GRBs
\citep{Woosley_1993, Kumar_&_Zhang_2015}. The inferred explosion
energy of these SNe is at least ten times larger
($E_{\rm HN} > 10^{52}$ erg) than that of typical SNe (see,
e.g. \citealt{Woosley_&_Bloom_2006}), which is why they are referred
in the literature as hypernovae (HNe, \citealt{Paczynski_1998,
  Iwamoto_1998}).  However the associated GRBs display a large variety
of isotropic energies.  A good number of the confirmed events have
GRBs classified as low luminosity bursts (llGRBs), because they show
equivalent isotropic luminosities in the range
$L_{\rm \gamma, iso}\sim 10^{46}$--$10^{48}\,$erg\,s$^{-1}$ (to be
compared with typical values of
$L_{\rm \gamma, iso}\sim 10^{50}$--$10^{53}\,$erg\,s$^{-1}$ for LGRBs;
see, e.g. \citealt{Hjorth_2013}). In addition to being less luminous
than canonical long GRBs, and despite some of them showing very long
durations ($\lesssim 1000$ s), llGRBs also seem to be less energetic
than the latter (holding equivalent isotropic energies
$E_{\rm \gamma, iso}\sim 10^{48}$--$\text{few}\,\times10^{49}\,$erg;
i.e. two to three orders of magnitude smaller than canonical GRBs),
are softer ($E_{\rm p} < 100$ keV), display relatively smooth
(non-variable) light curve (LC), and show no evidence for a
high-energy power-law tail \citep[e.g.][]{Bromberg_Nakar_Piran_2011}.

The atypical prompt emission, the origin of X-ray blackbody (BB)
components, and the unusual X-ray afterglow shown in many GRB/SN
associations, especially in the case of llGRBs (e.g. llGRBs 060218;
\citealt{Campana_etal_2006Natur}, 100316D; \citealt{Starling_2011,
  Cano_2011}, although not restricted to them; see
e.g. \citealt{Page_etal_2011MNRAS.416.2078} for GRB 090618) are
difficult to fit in terms of standard GRB theory.  In light of the
current observational data, it is still unclear whether progenitors of
llGRBs are the same as that of cosmological GRBs or whether these
super-long llGRBs are members of a low-luminosity end of a continuum
of collapsar explosions or, maybe, a different stellar endpoint.
Answering these questions has important implications for high-mass
stellar evolution, the connection between SNe and GRBs, and the
low-energy limits of GRB physics, especially taking into account that
llGRBs are likely more frequent than cosmological GRBs
\citep{Coward_2005, Cobb_2006, Pian_2006, Soderberg_2006a, Liang_2007,
  Guetta_&_Della_Valle_2007, Fan_et_al_2011}.  In particular,
\cite{Soderberg_2006a} calculated that the rate of SNe Ic-bl is about
the same as that of llGRBs, implying that llGRBs cannot be
significantly beamed. Indeed, it is likely that llGRBs are even
isotropic and the result of ``failed'' jets
\citep[e.g.][]{Bromberg_Nakar_Piran_2011}, i.e. jets which become
partly choked in the stellar envelope.  Due to the lack of prompt
observations in GRB/SN detection it is poorly understood how an
outgoing shock (relativistic or not) punches out its progenitor
star. In the collapsar model a relativistic jet must successfully
break out of its progenitor in order to produce a canonical
LGRB. Before the typical non-thermal radiation associated to the flash
in $\gamma$/X-rays caused by the breakout, the SN shock breakout may
produce some sort of thermal signal \citep*{Campana_etal_2006Natur,
  Waxman_Meszaros_Campana_2007, Soderberg_etal_2008Natur,
  Nakar_Sari_2012ApJ...747...88, Nakar_2015,
  Irwin_&_Chevalier_2016}. Observations in the last ten years are
pointing towards this direction. In particular the llGRB 060218,
associated to SN 2006aj, shows a thermal component in X-rays which
cools as it moves to optical frequencies
\citep{Campana_etal_2006Natur}. Based on the analysis of such
observations, the authors argue that the BB emission may arise from a
SN shock wave breaking out the extended wind surrounding a WR star.
%
%
More recently, \cite{Nakar_2015} has suggested that both canonical
LGRBs and llGRBs may have a similar origin. The similar properties
inferred from the associated SNe in both kinds of bursts suggest the
existence of similar progenitors with modified environments. The
optical LCs of the llGRB 060218, with a two peak structure, suggest
the presence of an extended low-mass envelope at a distance of
$10^{13}$--$10^{14}$\,cm that could also explain the observable
differences at high energies (see also
\citealt{Irwin_&_Chevalier_2016}). This envelope would brake the
incipient relativistic jet and would dissipate part of the energy
choking the jet that would emerge at much lower speeds without
producing a typical burst signal. Similar qualitative conclusions have
been obtained for the GRB 100316D/SN 2010bh \citep{Starling_2011} and
for the GRB 140606B /SN iPTF14bfu \citep{Singer_2015,Cano_2015}.  The
wide GRB energy ranges observed could be explained on the basis of
progenitor-dependent wind properties that might give rise to jets with
different conditions, with the relativistic ones as those that would
produce the most energetic bursts. Low luminosity GRBs may be
generated by mildly relativistic or even almost `failed' jets (see
above).

Our main goal here is to show that standard, low-metallicity stellar
progenitors of LGRBs may not produce relativistic jets of sufficiently
low luminosity to account for the least luminous end of the
distribution of llGRBs, unless either the produced jets undergo a
substantial broadening of their opening angle after they break out of
the surface of the stellar progenitor, or the radiative efficiency in
the gamma ray band is really small. Indeed, we will show analytically
(Sec.\,\ref{sec:threshold}) the existence of a luminosity threshold
below which, jet injection conditions are not well posed, since jets
must be supersonic with respect to the reference frame attached to the
progenitor star. This finding should be added to the previous criteria
for the production of LGRBs found by \cite{Bromberg_Nakar_Piran_2011},
inasmuch as the threshold we obtain is not set by the time over which
the central engine is efficiently pumping energy into a relativistic
outflow compared to the jet crossing time of the stellar progenitor.
In order to check the validity of our analytic conclusions, we have
developed a detailled numerical model (Sec.\,\ref{sec:model}) that
encompasses all the elements which are relevant to set the former
threshold, i.e. the structure of the massive stellar progenitor and
its circumstellar medium (a WR surrounded by an inflated envelope),
and the possible existence of a SN shock launched in addition to the
injection of a relativistic jet. Following the ideas of
\cite{Nakar_2015}, in this paper we work under the hypothesis that the
stellar progenitor is the same for canonical LGRBs, llGRBs and SN Ic
explosions.  For such reason we assume that a relativistic jet and a
SN ejecta can form inside of the stellar progenitor. Which outflow
forms first is still unclear, although numerical simulations suggest
that the SN would form first
\citep{Obergaulinger_Aloy__MNRASL__2017}. In this work, we consider
all possibilities.  In Section\,\ref{sec:simulations}, we contrast the
analytical results with simulations and we study the propagation of
relativistic jets within a WR star which may (or may not) have formed
a previous SN shock wave.  In Section\,\ref{sec:conclusions}, we
discuss our results in view of the existence of a luminosity threshold
constraining the generation of low-luminosity transients in potential
stellar progenitors of LGRBs/llGRBs.

\section{Luminosity threshold for jet injection and llGRBs production}
\label{sec:threshold}
%
A GRB jet can be launched inside of a massive star if a suitable
physical mechanism, which so far is not totally understood, extracts
sufficient energy from the central engine. The jet will propagate
subrelativistically within the star and eventually will break out of
the star relativistically after a time $t_{\rm b}$ (e.g.
\citealt{Aloy_etal_ApJL_2000__Collapsar}), before giving rise to the
$\gamma$-ray emission. To do so the central engine must be active a
time $t_{\rm e} > t_{\rm b}$ to compensate for the energy dissipated
in the jet in its way through the star.  Many numerical studies have
been carried out investigating the propagation of relativistic jets in
collapsars, either in 2D
\citep[e.g.][]{MacFadyen_Woosley_ApJ_1999__Collapsar,
  Aloy_etal_ApJL_2000__Collapsar, MWH_2001ApJ,
  Zhang_2003ApJ...586..356Z, Zhang_2004ApJ...608..365Z,
  Mizuta_etal_2006ApJ...651..960M, Morsony_2007,
  Morsony_2010,Lazzati_2010, Mizuta_Aloy_2009, Lazzati_2009,
  Nagakura_etal_2011ApJ...731...80N,Lazzati_2013,Lopez-Camara_2014}
and also in 3D \citep[e.g.][]{Zhang_2003ApJ...586..356Z,
  Lopez_Camara_2013,Ito_2015,Lopez-Camara_2016ApJ...826..180,Bromberg_2016MNRAS.456.1739}.

Should the progenitors of both LGRBs and llGRBs be the same
\citep{Nakar_2015}, it would be natural to assume that llGRBs
correspond to the low luminosity tail of LGRBs. However, relativistic
jets injected with a luminosity below a threshold that we may estimate
analytically, fail to penetrate the stellar envelope and, hence, they
won't produce a llGRB.  We shall show in this section that the latter
luminosity threshold is too large to explain events with luminosities
$\lesssim 10^{48}\,$erg\,s$^{-1}$ in standard LGRB progenitors.

Let us assume that a relativistic jet has already been formed and that
it has developed some fiducial conditions at radial distances
sufficiently much larger than the pre-collapsed iron core ($\gtrsim
10^{9}$\,cm), e.g. a well defined cross-sectional injection radius,
$R_{\rm j}$, a relatively small half-opening angle $\theta_{\rm j}\ll
1$ and a bulk Lorentz factor $\Gamma_{\rm j}=(1-v_{\rm j}^2)^{-1/2}>1$
(where $v_{\rm j}$ is the velocity of the beam). Under these
conditions, the complex problem of jet formation in the vicinity of
the central engine can be replaced by the injection of a relativistic
flow through a relatively narrow nozzle. The jet injection conditions
must guarantee that the outflow is supersonic with respect to the
external medium (otherwise, the mathematical injection problem is not
well-posed). That means that at the injection point the velocity of
the jet's head, $v_{\rm h}$, must be larger than the speed of sound of
the medium, $c_{\rm s,a}$, i.e.
\begin{equation}
  v_{\rm h} > c_{\rm s,a}.
\label{eq:supersonic}
\end{equation}
If, for simplicity, we neglect the gravitational pull of the star and
assume that the stellar matter is at rest, the velocity of the jet's
head can be approximated by
\citep{Marti_1997ApJ...479..151,Matzner_2003,Bromberg_2011}
\begin{equation}
v_{\rm h} = \frac{v_{\rm j}}{1 + \tilde{L}^{-1/2}}\, ,
\label{eq:headvelocity}
\end{equation}
where  
\begin{equation}
\tilde{L} \equiv \frac{\rho_{\rm j} h_{\rm j} \Gamma_{\rm j}^2}{\rho_{\rm a}} \simeq \frac{L_{\rm j}}{S_{\rm j} \rho_{\rm a} c^3}\, ,
\label{eq:Ltilde}
\end{equation}
where $L_{\rm j}$ is the jet luminosity, $h_{\rm j}$ is the specific
enthalpy of the jet and
\begin{equation}
S_{\rm j}= \pi R_{\rm j}^2 \sin{\theta_{\rm j}}^2 \approx \pi R_{\rm j}^2 \theta_{\rm j}^2
\label{eq:Sj}
\end{equation} 
is the jet's cross section. The indices `j' and `a' refer to
properties of the jet and the ambient medium, respectively. Since the
stellar density is much larger than the jet density, i.e. $\rho_{\rm
  j}/\rho_{\rm a}\ll 1$ and for typical GRB jet conditions the
asymptotic Lorentz factor $\Gamma_\infty:=h_{\rm j}\Gamma_{\rm j}>100$
and $1<h_{\rm j} \gtrsim 20$ (the upper bound would correspond to a
rather hot outflow and is only given for reference) we expect,
$\tilde{L} \ll 1$. Then, using Eq.\,(\ref{eq:supersonic}) we obtain
\begin{equation}
\frac{v_{\rm h}}{c_{\rm s,a}} \simeq \frac{\tilde{L}^{1/2}}{c_{\rm s,a}} > 1
\label{eq:supersonic-condition}
\end{equation}
From this condition, and using Eq.\,(\ref{eq:soundspeedTM}) and
(\ref{eq:Sj}), we estimate the minimum luminosity that a jet with a
half-opening angle $\theta_{\rm j}$ must have at the injection point,
$R_{\rm j} = R_{\rm inj,j}$, namely, $L_{\rm j} \gtrsim \Ljth$, where
\begin{equation}
\begin{split}
  \Ljth := &\, 1.4 \times10^{49} \\
  & \times \left(\frac{R_{\rm j}}{2 \times 10^9 \,\text{cm}}\right)^2
  \left(\frac{\theta_{\rm j} }{2^\circ}\right)^2 \left(\frac{p_{\rm
        a}}{1.8\times10^{22}\,\text{erg\,cm$^{-3}$}}\right)\,\text{erg
    s$^{-1}$}\, ,
\end{split}
\label{eq:Lj}
\end{equation}
where we have scaled the ambient medium pressure at $R_{\rm inj, j}$
employing typical values of the stellar progenitor 35OC (see
Sec.\,\ref{sec:progenitor}) at $R_{\rm inj, j}$. Also, we have taken
an injection half-opening angle ($\theta_{\rm j}=2^\circ$) as small as
possible to minimize the obtained luminosity threshold but, at the
same time, compatible with a hydrodynamic or MHD jet generation. To
convert the previous jet intrinsic luminosity into an equivalent
isotropic quantity we employ
$L_{\rm iso,j} = 2 L_{\rm j} / (1 - \cos{\theta}) \simeq 4 L_{\rm j} /
\theta^2$ where $\theta = \theta_j$ at the injection point. Likewise,
if this luminosity is maintained for an injection time
$t_{\rm inj,j}$, after which it is shut down progressively (see
Sec.\,\ref{sec:jets}), we obtain an equivalent isotropic energy from
$E_{\rm iso,j} = L_{\rm iso,j} \times (4/3)\, t_{\rm inj,j}$; where
the factor $4/3$ takes into account the shut down phase of the
injection.  Thus, at the injection point:
\begin{equation}
L_{\rm iso,j} \gtrsim 4.5\times10^{52} \, \left(\frac{R_{\rm j}}{2\times10^9 \,\text{cm}}\right)^2  
\left(\frac{p_{\rm a}}{1.8\times10^{22}\,\text{erg\,cm$^{-3}$}}\right)\,\text{erg s$^{-1}$}\, ,
\label{eq:luminosity-threshold}
\end{equation}  

and
\begin{equation}
\begin{split}
  E_{\rm iso,j} \gtrsim 1.2\times10^{54} \, & \left(\frac{R_{\rm
        j}}{2\times 10^9 \,\text{cm}}\right)^2
\\ 
\times & \left(\frac{p_{\rm a}}{1.8\times10^{22}\,\text{erg\,cm$^{-3}$}}\right) \left(\frac{t_{\rm inj,j}}{20\,\text{s}}\right)\,\text{erg}\, .
\end{split}
\label{eq:energy-threshold}
\end{equation}   

We note that the former estimates of the isotropic jet luminosity and
energy have been made assuming that the value of the outflow
half-opening angle is the same asymptotically as it is
initially. Nevertheless, according to \cite{Mizuta_&_Ioka_2013}, after
the jet breakout the opening angle of the jet becomes $\theta_{\rm
  BO}\simeq 1/(5\Gamma_{\rm j})$. The value of the beam bulk Lorentz
factor at the injection point shall be rather moderate, since it is
very close to the stagnation point from where the outflow is
launched. Thus, the acceleration of the beam has not finished yet, and
most of it may happen while the jet is travelling through the outer
stellar layers and after the jet break out
\citep[e.g.][]{Aloy_etal_ApJL_2000__Collapsar,Mizuta_&_Ioka_2013}. In
practical terms, $\Gamma_j\lesssim 10$ are likely values for the jet
injection Lorentz factor. For the purposes of this estimate, we use
$\Gamma_{\rm j}=5$ (see Sec.\,\ref{sec:jets}) and, therefore, we would
obtain $\theta_{\rm B0}\simeq 2.3^\circ\simeq \theta_j$, so that the
previous thresholds for both the isotropic luminosity
(Eq.\,\ref{eq:luminosity-threshold}) and for the isotropic
energy (Eq.\,\ref{eq:energy-threshold}) of the jet would remain at
breakout since=
\begin{equation}
L_{\rm iso, BO} \simeq 4 L_{\rm j} / \theta_{\rm BO}^2 \simeq L_{\rm iso,j}\, ,
\label{eq:BOluminosity}
\end{equation}
and also $E_{\rm iso,BO} \simeq E_{\rm iso,j}$.

Hence, if the results of \cite{Mizuta_&_Ioka_2013} hold, we do not
expect any jets that manage to propagate supersonically in the
pre-supernova model 35OC to fall into the category of llGRBs, unless
the radiative efficiency in the $\gamma-$ray band,
$\epsilon_\gamma=L_{\rm iso,\gamma} / L_{\rm iso}$, is very small
($\epsilon_\gamma\lesssim 10^{-4}$). However, the breakout angle can
be significantly larger than predicted by \cite{Mizuta_&_Ioka_2013}
for jets injected in the stellar progenitor model 16TI of
\cite{Woosley_&_Heger_2006}, especially if the jet luminosity is very
close to the threshold we have found (see
Sec.\,\ref{sec:simulations}).  Indeed, we expect $\theta_{\rm BO} \sim
a / \Gamma_{\rm j}$, with $0.5< a \lesssim 3$. This larger breakout
angle translates into the following constrains on the isotropic
luminosity and energy:
\begin{equation}
\begin{split}
L_{\rm \gamma, iso, BO} \gtrsim \, 1.5\times10^{48} \, &
\left(\frac{\epsilon_\gamma}{0.01} \right) \, 
\left(\frac{R_{\rm j}}{2\times 10^9 \,\text{cm}}\right)^2 
\left(\frac{\theta_{\rm j}}{2^\circ} \frac{35^\circ}{\theta_{\rm BO}} \right)^2 \\
\times & \left(\frac{p_{\rm a}}{1.8\times10^{22}\,\text{erg\,cm$^{-3}$}}\right)\,\text{erg s$^{-1}$}\, ,
\end{split}
\label{eq:luminosity-threshold2}
\end{equation}  
and 
\begin{equation}
\begin{split}
E_{\rm \gamma, iso, BO} \gtrsim \, 3.9\times10^{49} \, & 
\left(\frac{\epsilon_\gamma}{0.01} \right) 
\left(\frac{R_{\rm j}}{2\times 10^9 \,\text{cm}}\right)^2 
\left(\frac{\theta_{\rm j}}{2^\circ} \frac{35^\circ}{\theta_{\rm BO}} \right)^2 \\
\times &
\left(\frac{p_{\rm a}}{1.8\times10^{22}\,\text{erg\,cm$^{-3}$}}\right)
\left(\frac{t_{\rm inj,j}}{20\,\text{s}}\right)\,
\text{erg} .
\end{split}
\label{eq:energy-threshold2}
\end{equation}  
Therefore, it is possible to host in model 35OC jets, which would have
isotropic equivalent luminosities consistent with the upper bound of
llGRBs if, as suggested by our models (see next
Sec.\,\ref{sec:simulations}), the breakout opening
angle grows well above the expectations of \cite{Mizuta_&_Ioka_2013},
and if the acceptable radiative efficiency is $\epsilon_\gamma\lesssim
1\%$.

\begin{figure}
\includegraphics[width=0.49\textwidth]{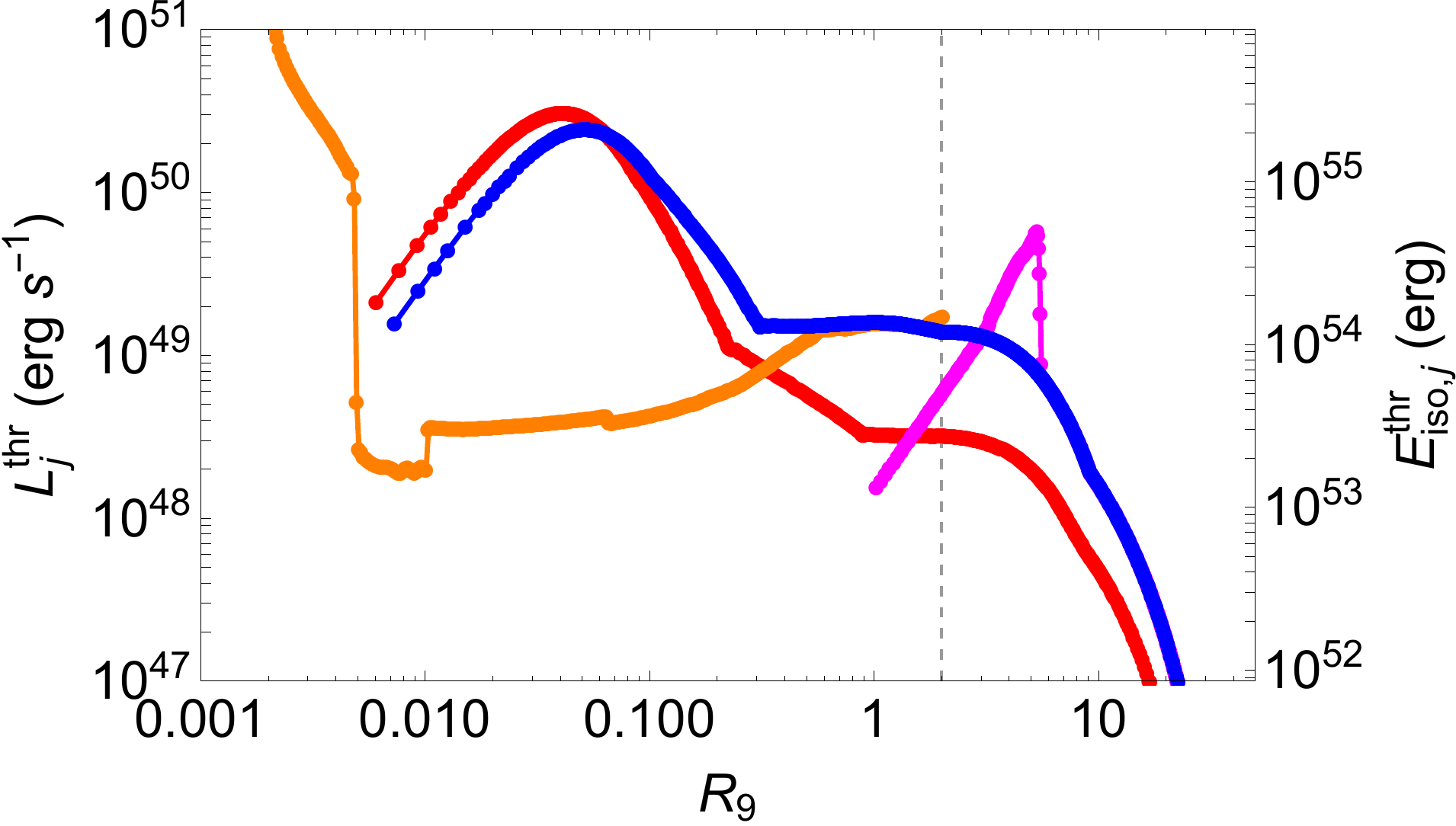}
\caption{Intrinsic luminosity threshold (Eq.\,\ref{eq:Lj}; left scale)
  and equivalent isotropic energy threshold
  (Eq.\,\ref{eq:energy-threshold}; right scale) for different
  progenitor models: pre-supernova models 35OC (blue) and 16TI (red)
  of \citealt{Woosley_&_Heger_2006}, and initial model of the series
  J3 (Tab.\,\ref{tab:cjet-models}) of this work (magenta). The latter
  models include a SN shock wave set up with an energy
  $E_{\rm SN}=10^{52}$\,erg.  We also include data computed for the
  density profile of one of the pre-supernova models of
  \citealt{Obergaulinger_Aloy__MNRASL__2017}, corresponding at the
  moment of BH formation after the core collapse of model 35OC
  (orange). The vertical dashed line shows the position of
  $R_{\rm inj} (\emph{Jet})$. The radius is expressed in units of
  $10^9$\,cm, $R=R_9 10^9$\,cm. Each of the models displayed in the
  figure is computed on a different radial grid. The stellar models
  16TI and 35OC on the original radial grid of
  \citealt{Woosley_&_Heger_2006}, while models J3 are computed on a
  grid where the central part of the 35OC progenitor ($R_9<1$) is
  excised and that extends to the stelar radius $R_\ast$
  (Sec.\,\ref{sec:model}). Finally, the model evolved to the brink of
  BH collapse is computed on a grid spaning $R_9<2.5$. In all cases,
  we assume that jets are injected with a half-opening angle
  $\theta_{\rm j}=2^\circ$. Note that we have assumed that the
  breakout half-opening angle equals $\theta_{\rm j}$ in order to
  compute $E_{\rm iso,j}^{\rm thr}$.}
\label{fig:energy-threshold}
\end{figure}

The jet equivalent isotropic energy threshold
(Eq.\,\ref{eq:energy-threshold}) depends upon three factors. One of
them, $t_{\rm inj}$, is an intrinsic jet property, which cannot be
much shorter than $\sim 20\,$s, if we aim to produce jets which
release $\gamma-$radiation over time scales compatible with llGRBs or
LGRBs. The ambient medium pressure at $R_{\rm j}$ may be different
depending on the radial distance at which we inject the jet. In order
to assess the robustness of the criterion found, we have computed the
equivalent isotropic energy as a function of the radius
(Eq.\,\ref{eq:energy-threshold}) for a jet to form with an intrinsic
luminosity right at the threshold found in Eq.\,(\ref{eq:Lj}). For
this purpose, we assume that the outflow half-opening angle is kept
approximately constant and equal to $\theta_{\rm j}$.  The results for
two different progenitors and different variations of the physical
conditions inside them are shown in Fig.\,\ref{fig:energy-threshold}.
The blue line corresponding to progenitor model 35OC shows that jets
with $E_{\rm iso,j}^{\rm thr}\lesssim 10^{54}$\,erg cannot be launched
unless the injection radius is $R_{\rm j}\gtrsim 6\times
10^9$\,cm. Beyond this distance, and up to $R_{\rm j}\simeq 2\times
10^{10}\,$cm only jets with $E_{\rm iso,j}^{\rm thr}\gtrsim
10^{52}$\,erg can be launched. The reduction in the luminosity
threshold for $R_{\rm j}\gtrsim 6\times 10^9$\,cm is produced by the
quick decline of the pressure (and hence of the $c_{\rm s,a}$) in the
envelope of the stellar progenitor. But even if jet fiducial
conditions would set in the stellar envelope, $L_{\rm iso,j}$ would be
only marginally compatible with the luminosities observed for the most
powerful llGRBs.

In order to reduce the luminosity threshold found in
Eq.\,(\ref{eq:Lj}), we may seek ways to reduce the pressure or the
density of the stellar progenitor model. One possibility to obtain the
sought effect is including a SN shock wave that propagates from the
outer stellar core towards the stellar surface. We mimic the global
effect of such a SN shock wave (see Sec.\,\ref{sec:SNejecta}) setting
it up with a ``piston'' mechanism assuming spherical symmetry and
injecting the SN in our stellar progenitor model at a radial distance
$R_0=10^9\,$cm (App.\,\ref{sec:piston}). The SN shock wave possesses
an energy, $E_{\rm SN}=10^{52}\,$erg, and we let it propagate inside
of the 35OC stellar model up to a distance of
$\sim 5\times 10^{9}\,$cm by the time we compute the local values of
$\Ljth$ and of $E_{\rm iso,j}^{\rm thr}$ that are displayed in
Fig.\,\ref{fig:energy-threshold} (magenta line).  As a result of the
passage of the SN shock the density in its wake decreases
substantially and lowers significantly the luminosity threshold (by
about one order of magnitude) to launch a jet up to
$\sim 3\times 10^9\,$cm. Beyond that radial distance, the lines
corresponding to the thresholds in the original 35OC model (blue line
in Fig.\,\ref{fig:energy-threshold}) and the SN exploding model
(magenta line) cross.  Remarkably, if we inject a jet with a delay of
a few seconds after the passage of the SN shock at a distance of
$\sim 10^9\,$cm, the jet's head will be initially supersonic even if
it is launched with a luminosity of about one order of magnitude
smaller than in the original progenitor (where the reduced luminosity
would not allow for a successful jet initiation). We note that, fixed
the injection radius, e.g. at $R_{\rm j}\sim 10^{9}\,$cm, the
reduction in the luminosity injection threshold is proportional to the
time delay between the passage of the SN shock by $r=R_{\rm j}$ and
the injection of the jet; the longer the delay, the lower the
luminosity threshold to initiate the jet. Nevertheless, we do not
expect that the time delays between the SN shock and the jet injection
be much longer than a few seconds (see Sec.\,\ref{sec:jets}) and,
thereby, the reduction of the luminosity threshold should not be much
larger than depicted in Fig.\,\ref{fig:energy-threshold}. This means
that, even if the SN shock precedes the jet injection, the expected
equivalent isotropic luminosity of a {\em potentially} succesful jet
would not be low enough to account for most of the lower luminosity
llGRBs. We further note that the SN shock compresses the stellar
matter and, as a result, the luminosity threshold is larger than that
corresponding to the unperturbed stellar progenitor in a fraction of
the shocked region, namely, in the region $3\lesssim R_9 \lesssim 5$;
Fig.\,\ref{fig:energy-threshold}. Should the jet be initiated below
this region, it would interact with the SN shocked matter and,
depending on its intrinsic luminosity, it could be choked inside the
start (see Sec.\ref{sec:jets}).

Admittedly, the set up of the SN shock wave by means of a piston
mechanism may be an oversimplification of the (much) more involved
process of SN shock generation mechanism. Thus, we have also evaluated
the jet luminosity threshold and the corresponding value of
$E_{\rm iso,j}^{\rm thr}$ as a function of radius for a more realistic
model. Figure\,\ref{fig:energy-threshold} (orange line) also includes
the profile of a pre-supernova model of
\citeauthor{Obergaulinger_Aloy__MNRASL__2017}
(\citeyear{Obergaulinger_Aloy__MNRASL__2017}; orange line; model OA17
hereafter), evolved from the onset of core collapse in the progenitor
35OC until BH formation. We chose a non-exploding model for this
comparison in order to study how much the decrease of the gas density
in the core due to the accretion of matter onto the proto-neutron star
lowers the threshold for jet production.  By the end of the
simulation, the layers at the distance of our injection location are
still unaffected by the dynamics of the core collapse and, thus, there
is effectively no difference in the luminosity threshold for jet
initiation with respect to our reference 35OC model for $R_9\ge
1$. The tiny differences observed close to $R_{\rm inj}(\emph{Jet})$
(identified by the vertical dashed line in the figure) are due to the
influence of the boundary conditions in the simulations of
\citeauthor{Obergaulinger_Aloy__MNRASL__2017}, whose numerical grid
spans the region contained up to $R_9 =2.5$ (since they are mostly
interested in the dynamics of the central engine). We remark that in
the more realistic initial profile of model OA17, their fiducial jet
injection conditions are set deeper inside than in our model (e.g. at
$0.1R_9$), the luminosity threshold decreases by a factor $\sim 5$
with respect to our reference 35OC model initiating the jet at
$R_{\rm inj}(\emph{Jet})$. Nonetheless, this reduction is still
insufficient to drive a supersonic jet inside the grid with a
luminosity compatible with that of llGRBs.

In order to explore the progenitor dependence of the luminosity
threshold, we depict the values of $\Ljth$ and of
$E_{\rm iso,j}^{\rm thr}$ for another GRB progenitor candidate, the
pre-supernova model 16TI of \cite{Woosley_&_Heger_2006}. For this
model both the luminosity and the energy threshold is smaller than
that of the model 35OC above $10^8$\,cm, e.g. by a factor 5 around
$10^9$\,cm. However, as in all the previous cases, the reduction in
the luminosity threshold does not suffice for the purpose of
initiating a llGRB jet, if the jet breakout opening angle is similar
to the jet injection half-opening angle and the radiative efficiency
is not as low as $\epsilon_\gamma\simeq 10^{-3}$ (see the discussion
below Eq.\,(\ref{eq:energy-threshold2})).

\section{The numerical model}
\label{sec:model}

In order to test the analytic results for the existence of a
luminosity threshold obtained in Sec.\,\ref{sec:threshold}, we have
set up a number of numerical models. These models include the massive
stelar progenitor 35OC (Sec.\,\ref{sec:progenitor}) endowed with an
extended envelope characteristic of WR stars
(Sec.\,\ref{sec:envelope}). Furthermore, we may include the effects of
a parametrized SN shock wave (Sec.\,\ref{sec:SNejecta}) modifying the
profile of the stellar progenitor in which relativistic jets of
different luminosities (close to the thresholds found in
Sec.\,\ref{sec:threshold}) are injected (Sec.\,\ref{sec:jets}).  The
simulations have been performed with the relativistic hydrodynamics
code MRGENESIS \citep{Aloy_1999ApJS,
  Leismann_etal__2005__ApJ__RMHD-Jets, Mimica_etal_2009ApJ}. We
modelled the outflows, the stellar gas and the circumstellar medium
using the \emph{TM} equation of state \citep{Mignone_2005ApJS,
  Mignone_&_McKinney_2007}, which, while not accounting for the
detailed chemical composition of the gas and for radiation, is a valid
approximation in our case.  For resolving the large gradients between
the outflows and the stellar profile, we have used the third-order
spatial reconstruction scheme PPM
\citep{Colella_Woodward_1984JCoPh..54..174}.  In order to maintain the
hydrostatic equilibrium of the star, we considered its self-gravity by
including a Newtonian gravitational potential in MRGENESIS
(App.\,\ref{sec:gravity}).

The numerical evolution of our models has been divided in two
steps. Firstly, we perform 1D simulations of SN ejecta propagating
into the progenitor star (Sec.\,\ref{sec:SNejecta}). Secondly,
we perform 2D simulations of relativistic jets propagating into (1)
the medium left behind by the SN ejecta or (2) the progenitor star
without modification (Sec.\,\ref{sec:jets}).
\begin{figure}
\includegraphics[width=0.47\textwidth]{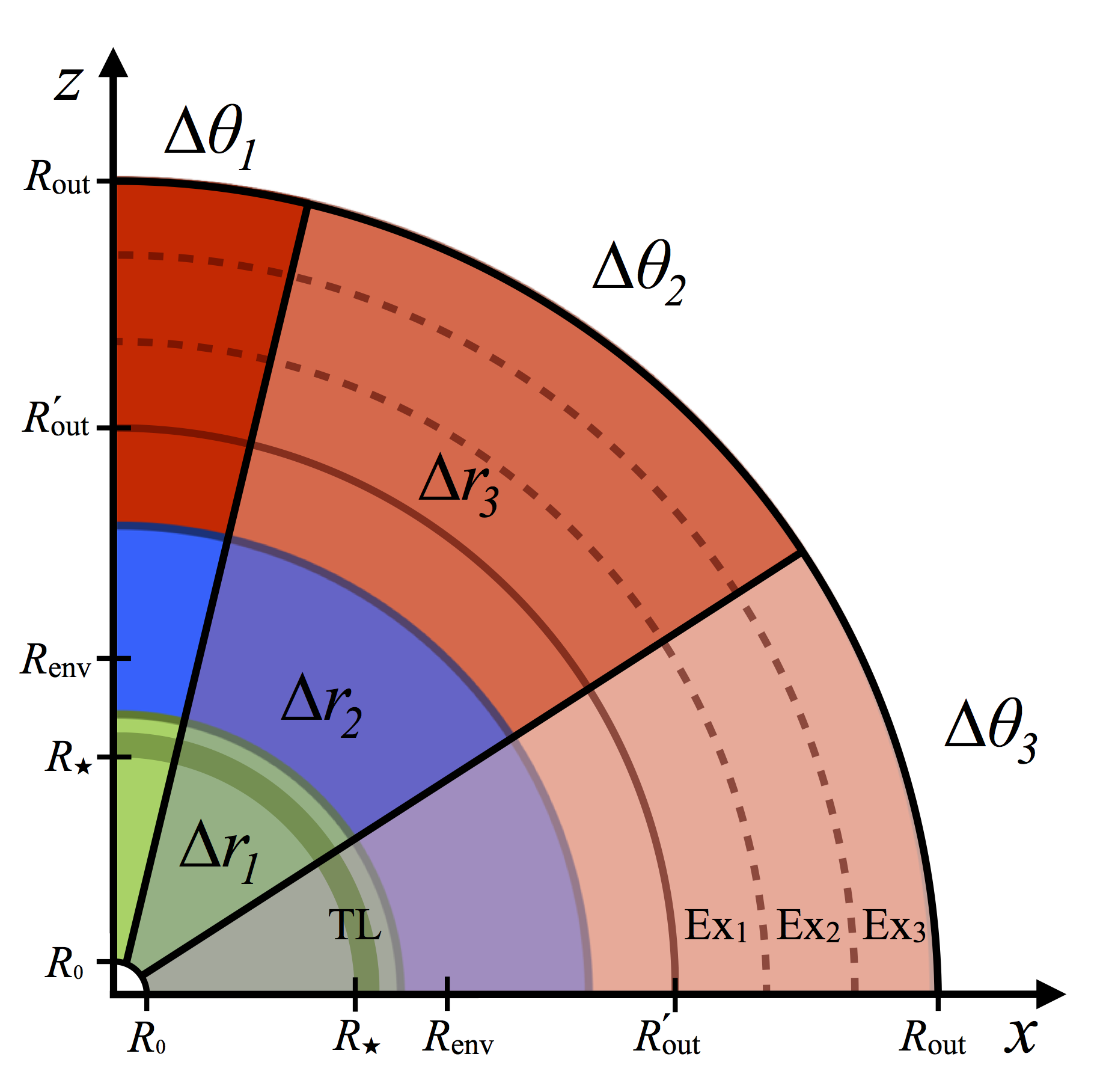}
\caption{Sketch of the numerical grid. We use three subgrids
    with different levels of refinement along both radial (green, blue
    and red) and angular directions (dark to lighter colours). The
    stellar surface ($R_\star$), the transition layer (TL; dark green
    annulus; included in the first level of refinement of the radial
    grid), the envelope extension ($R_{\rm env}$) and the the initial
    ($R'_{\rm out}$) and final length of the radial grid
    ($R_{\rm out}$) after the inclusion of three extra chunks (denoted
    as 'Ex') are marked in the figure. Note that the sketch
    particularly describes the grid setup in model SN. For jet
    simulations the sketch is equally valid but $R_0$ should be
    replaced by $R_{\rm inj,j}$. Furthermore, we stress that
    $R'_{\rm out}$ and $R_{\rm out}$ have different values and that
    only two extra chunks, with a different extension than those in
    model SH, are added to the grid in jet models. See the text for
    more details. Figure is not to scale.}
\label{fig:numerical-grid}
\end{figure}

The numerical resolution of the two-dimensional spherical grid is the
result of a trade-off between the computational cost and several
factors demanding spatially extended and fine grids.  On the one hand,
we need to resolve the small scales within the star to properly
resolve the jet/star interaction. On the other hand, we need to go to
larger length scales as we have to let the jet evolve up to very late
times. Obviously, in 1D we could employ much finer numerical
grids. Nevertheless, much larger resolutions are not viable in the
subsequent computational phase in 2D (Sec.\,\ref{sec:jets}). Our
choice has been to use a nested grid composed of three different
uniform-space subgrids (Fig.\,\ref{fig:numerical-grid}),
covering the range $[R_{\rm inj}, R'_{\rm out}] = [10^9, 2\times
10^{12}]\,$cm. The first radial subgrid has $n_{\rm r,1} = 1200$ zones
and covers the range $[R_{\rm inj}, 6.1\times 10^{10}\,$cm] (i.e. the
resolution is $\Delta r_1 = 5\times10^7$\,cm), while the second one
has $n_{\rm r,2} = 9400$ zones and covers the range $(6.1\times
10^{10}\,, 10^{12}$]\,cm ($\Delta r_2 = 10^8$\,cm) and the third one
has $n_{\rm r,3} = 2500$ zones and covers the range $(1, 2]\times
10^{12}$\,cm ($\Delta r_3 = 2\times10^8$\,cm).  Since our models need
to be run for a rather long time, the set up flows may eventually
reach the limit of the basic grid sketched above. When this happens,
we extend the computational domain in the radial direction. The
extension is done by adding chunks of external medium with the same
resolution as in the third level ($\Delta r_3$).  The injection
radius, $R_{\rm inj}$, for the different outflows we consider (the SN
ejecta and the jet) is different in each of the cases ($R_{\rm inj}
(\emph{SN}) = R_0 = 10^9$ cm and $R_{\rm inj} (\text{\emph{Jet}}) = 2
R_0 = 2 \times 10^9$ cm).

\subsection{The progenitor star}
\label{sec:progenitor}

\begin{figure}
\centering
\includegraphics[width=0.47\textwidth]{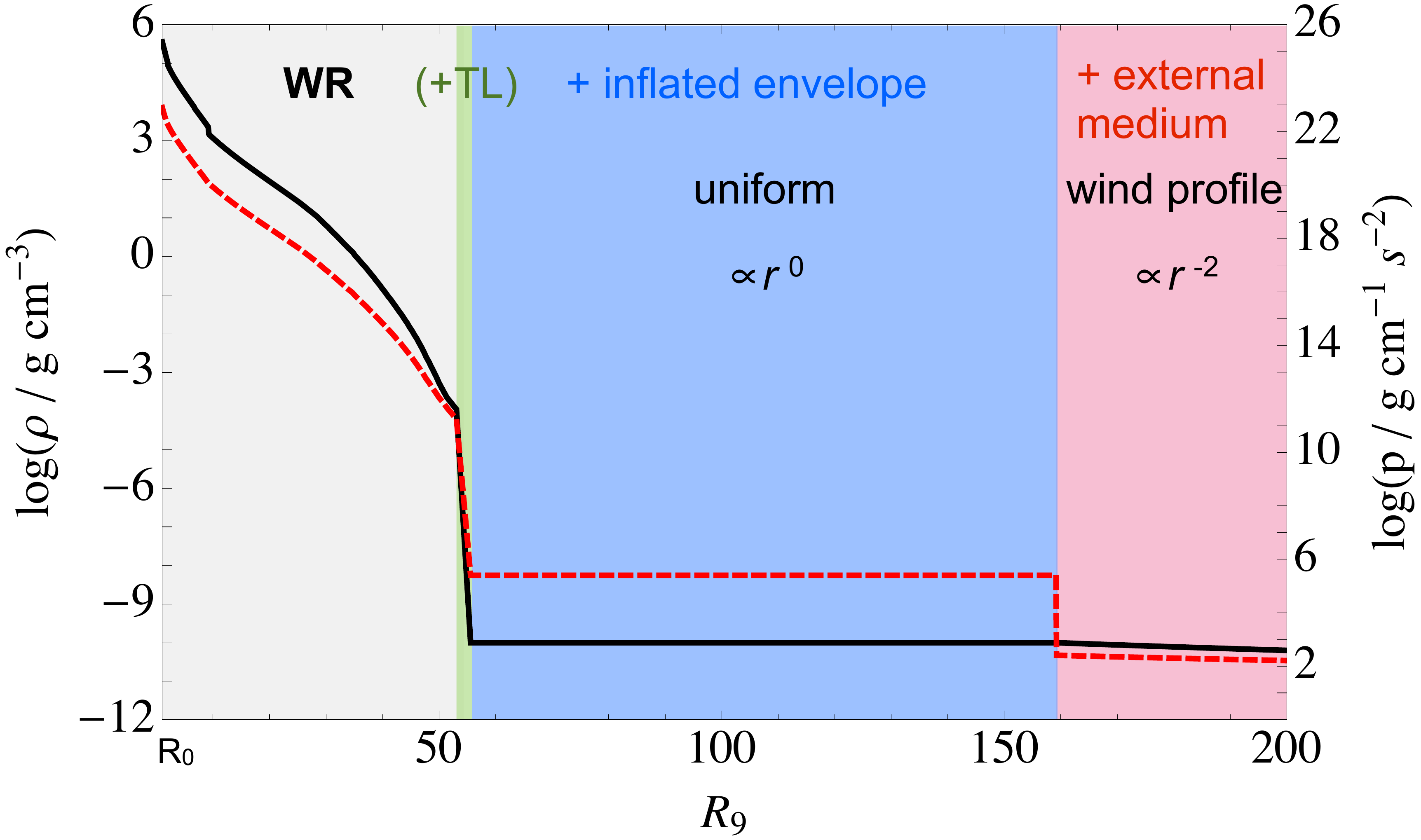}
\caption{Rest-mass density (black solid) and pressure (red
    dashed) profiles of model 35OC \citep[][the original star has a
    radius $R_9\simeq 52$]{Woosley_&_Heger_2006}, in addition to the
    inflated WR envelope (blue shaded background) and the wind-like
    external medium (pink shaded background). Between the WR stellar
    core and the inflated envelope there is a thin transition layer
    (TL; green shaded) set for numerical convenience.}
\label{fig:star-profile}
\end{figure}

As progenitor star, we have chosen the most massive star evolved by
\cite{Woosley_&_Heger_2006} that has enough rotational energy to be
considered a GRB candidate: the pre-supernova model 35OC. It
corresponds with a zero-age main-sequence (ZAMS) of $35 M_\odot$ that
reaches core collapse as a WR star with a final mass of
$28.07\, M_{\odot}$ and an iron core of $2.02 \, M_{\odot}$ with a
final stellar radius of $R_{\star} = 5.31 \times 10^{10}$\,cm.  From
this model we take rest-mass density (Fig.\,\ref{fig:star-profile}),
pressure and radial velocity profiles.  We ignore the angular velocity
profile for several reasons: (1) the rotational kinetic energy is very
small in the layers of the star beyond $10^9$\,cm, and (2) in order to
keep more easily the hydrodynamic equilibrium of the stellar model.
We have neither considered the chemical composition of the different
layers nor the explosive nuclear burning that may take place in the
shocks driven by the relativistic outflows.  The inner iron core is
excised up to $R_{\rm inj}$ and the remaining stellar progenitor is
mapped in our computational domain. Physically, the outer layers of
the star are in hydrostatic equilibrium and no causal connection
exists between them and the core on time scales smaller than the
free-fall time of each mass shell onto the core.  The excised masses
from the progenitor star beneath $R_{\rm 0}$ and $2 R_0$ are
$M_{\rm in} \approx 3.385 M_\odot$ and $\approx 6.055 M_\odot$,
respectively. We assume that these masses will collapse to form a PNS
first and, eventually, a BH \citep[e.g.][]{OConnor_&_Ott_2011,
  Cerda-Duran_2013,Obergaulinger_Aloy__MNRASL__2017}.

\subsection{Inflated envelope and external medium}
\label{sec:envelope}

Although the stellar-evolution calculations the model is based on
\citep{Woosley_&_Heger_2006} include all important processes
(e.g. rotation and magnetic fields), the approximations required for
following the evolution during the hydrostatic phases entail
limitations in several aspects. One of those is the mass loss due to
the intense winds driven by radiation pressure common to high-mass
stars, which modify the environment of the star and may generate
inflated envelopes or, in the most extreme cases, generate dense
shells around the star.  Mass loss is included in the models in a
parametrized way without accounting for the detailed structure of the
stellar winds beyond the surface of the star.  Consequently, we have
to set the profiles of density and temperature outside the star in
such a way that they are consistent both with typical mass loss rates
of this class of stars and with the properties of the GRB/SN
progenitors we want to model (see
e.g. \citealt{Crowther__2007__ARAA}).  Recent studies \citep*[see,
e.g.][]{Graefener_2012, Sanyal__2015__WR} consider that, instead of a
typical wind, WR stars may develop very dilute `inflated' envelopes
that can extend far beyond ($R_{\rm env} \sim 10^{12}$ cm) the
location of the core ($R_{\rm c} \lesssim 10^{11}$) and increase the
opacity, reconciling the large WR radii observed with the idea of
compact WR cores as GRB progenitors.  Based on this hypothesis, we
consider that an inflated envelope with a uniform rest-mass density,
$\rho_{\rm env} \gtrsim 10^{-10} \,$g\,cm$^{-3}$, and pressure,
$p_{\rm env} \sim 10^6$\,g cm$^{-1}$ s$^{-2}$, is generated at the
surface of the WR (see e.g. Fig.\,2 of \citealt{Graefener_2012} for
their most massive WR model of $23 M_\odot$). This envelope is not
contained in the stellar evolution model of
\cite{Woosley_&_Heger_2006}.  With these values both the rest-mass
density and pressure show strong jumps at the interface separating the
WR star and the envelope. Thus, for reasons of numerical stability, we
have implemented a transition layer consisting of 50 radial zones in
the layer $[R_\star, 5.56\times 10^{10}\,{\rm cm}]$ to smoothly
connect these two regions (Fig.\,\ref{fig:numerical-grid}).  The
radial extent of the inflated envelope,
$\Delta R_\mathrm{env}:=R_{\rm env}-R_\star$, was found by
\cite{Graefener_2012} to be a few stellar radii ($R_{\star}$), with
higher values for higher stellar masses.  Lacking detailed information
for model 35OC, our choice of
$\Delta R_\mathrm{env} = 2 R_{\star} \approx 1.1 \times 10^{11}$\,cm,
is motivated by the fact that a much larger value would make the
envelope optically thick, in conflict with the fact that stars that
end as fast rotating WR stars may become transparent to UV radiation
during the main-sequence evolution \citep{Szecsi_2015}.  From
$R_{\rm env}$ outwards we assume that both rest-mass density and
pressure follow a wind profile, i.e. $p$, $\rho \propto
r^{-2}$. Immediately outside of $R_{\rm env}$, i.e. in the interface
that separates the inflated envelope from the wind-like medium, we fix
$\rho_{\rm EM,0}=\rho_{\rm env}$ and
$p_{\rm EM,0} = 2.5 \times 10^5$\,g cm$^{-1}$ s$^{-2}$.  Though, the
wind is moving with a subrelativistic speed away from the star and the
envelope may be lossing mass, the speeds of both media are much
smaller than those the jets we set up in Sec.\,\ref{sec:jets}
develop. Thus, in practice, we assume that both the wind-like medium
and the envelope are at rest.  For simplicity, hereafter we will refer
to the wind-like external medium as just the external medium (EM).

\subsection{SN ejecta}
\label{sec:SNejecta}

Detailed models of core collapse of the pre-supernova model 35OC
\citep{Obergaulinger_Aloy__MNRASL__2017} show that a SN explosion may
develop within less than a second after core bounce. While the shock
wave propagates outwards, the conditions at the center might lead to
the launching of a GRB jet after a further delay of the order of a few
seconds (see further discussion in the next Section).

In order to incorporate in our model the dynamics generated by an
on-going SN explosion, we inject in our stellar progenitor at
$R_{\rm inj} (\emph{SN}) = R_0$ a synthetic SN ejecta driven \emph{ad
  hoc} by a {\em piston}-like mechanism (App.\,\ref{sec:piston}).
This injection radius has been chosen because it puts our inner
boundary well outside the iron core and the surrounding matter that
has fallen onto the hypermassive PNS by the time we start our
simulations \citep[see][]{Obergaulinger_Aloy__MNRASL__2017}.  We have
simplified our set up reducing the SN ejecta propagation to a 1D,
spherically symmetric problem. Certainly, we do not expect a SN shock
launched in a fast rotating stellar progenitor (as the one at hand) to
be perfectly isotropic, however, beyond the further simplified jet/SN
ejecta interaction, the 1D ejecta modelling reduces considerably the
computational cost associated to the calculation of its hydrodynamic
evolution as it travels inside of the stellar envelope.

  The initialization of the SN ejecta is done using a piston-like
  model similar to that of \cite{Rosenberg_&_Scheuer_73} or
  \cite{Gull_73}, since energy, $E_{\rm SN}$, and mass, $M_{\rm SN}$,
  are carried by an spherical flow, which enters the numerical grid
  through the inner boundary, at a constant rate, until
  $t_{\rm SN} = 1$\,s (see, Tab.\,\ref{tab:cjet-models}). The stellar
  potential sets a minimum required amount of energy to launch any
  outflow at a distance $R_0$. We have checked that the binding energy
  of the matter on the numerical domain is
  $\gtrsim 5\times10^{50}$\,erg, i.e. any successful SN must exceed
  this value, which is compatible with energies of typical SN. Thus,
  the total energy injected is $E_{\rm SN}=10^{52}\,$erg, which places
  this model in the HN realm (consistently with GRB/SN associations),
  while the total mass injected by the piston mechanism through the
  innermost radial boundary has been fixed to
  $M_{\rm SN} = 0.1 M_\odot$\footnote{This mass must not be confused
    with the mass of the SN ejecta, which is obviously much larger. It
    is only a practical way of setting up the properties of the piston
    mechanism.}. This mass is removed from the excised mass enclosed
  below $R_0$.  The reason for this operational procedure is to not
  modify the gravitational potential (and, hence, the equilibrium
  conditions) in the layers of the star beyond the SN shock. In case
  we do not apply this correction to $M_{\rm in}$, unwanted
  displacements of the outer progenitor mass shells (including the
  stellar surface) are generated (see App.\,\ref{sec:gravity}).  Once
  the constant injection phase is over, a quickly decaying mass and
  energy injection follows. After that, the inner boundary is open and
  copy conditions are set, allowing the inflow of material from the
  grid to the excised part. During this phase, we follow the evolution
  of $M_{\rm in}$, which is suitably updated to self consistently
  compute the gravitational potential.

 \begin{table*}
\centering
\caption[Models]{Summary of some of the parameters of all models:
  dimensionality (Dim),  true luminosity ($L$), isotropic equivalent energy ($E_{\rm iso}$),
  true energy ($E$), initial Lorentz factor ($\Gamma$), initial
  specific enthalpy ($h$), injection time ($t_{\rm inj}$),
  injection radius ($R_{\rm inj}$), inner mass below $R_{\rm inj}$
  ($M_{\rm in}$) and time delay with respect to the SN ($t_{\rm
    del}$). The true luminosity corresponds to $E_{\rm
    SN}/t_{\rm inj}$  and to  $L_{\rm j}$ for the SN and jet
  injection cases, respectively. $E_{\rm iso}$ is computed assuming
  that the half-opening angle at jet breakout of the stellar surface
  is approximately the same as that at injection ($\theta_{\rm
    BO}\simeq \theta_{\rm j}$).}
\begin{tabular}{|l|r|r|r|r|r|r|r|r|}
\hline Model 		     & SN 	        & J0a                        & J0b                        & J0c    	                    & J3a 			    & J3b                          &  J3c                        &  J3d\\\hline
Dim				     & 1D  	        & 2D                        &         2D 	         &         2D 		    & 2D 	      		    & 2D                          & 2D                          & 2D\\
$L$ (erg\,s$^{-1}$)        & $10^{52}$ &$1.1\times10^{49}$&$5.7\times10^{48}$&$2.9\times10^{48}$ &$1.1\times10^{49}$ &$5.7\times10^{48}$  &$2.9\times10^{48}$ & $1.1\times 10^{48}$ \\ 
$E_{\rm iso}$ (erg) 	     & $10^{52}$ & $10^{54}$              &$5.0\times 10^{53}$&$2.5\times 10^{53}$&   $10^{54}$    	    &$5.0 \times 10^{53}$&$2.5\times 10^{53}$& $10^{53}$ \\
$E$ (erg) 		             & $10^{52}$ & $3\times10^{50}$ &$1.5\times 10^{50}$&$7.6\times 10^{49}$&$3\times   10^{50}$ &$1.5\times10^{50}$  &$7.6\times 10^{49}$&$3\times 10^{49}$ \\ 
$\Gamma$ 		     & $1.0315$ & 5                          & 5                            &               5              & 5 		            & 5                             & 5                            &  5 \\
$h$ 				     & $1.0237$ & 20                        &20                           & 20                          &20 			    & 20                           &20                           &  20 \\
$t_{\rm inj}$ (s) 		     & 1 	        & 20                        & 20                          & 20                          &20 			    & 20                           &20                           &  20 \\
$R_{\rm inj}$ 	    	     & $R_0$      & $2R_0$                 &$2R_0$                    & $2R_0$                   &$2R_0$ 		    & $2R_0$                    &$2R_0$                   & $2R_0$\\
$M_{\rm in}$ ($M_\odot$) & $3.385$   & $6.055$               & $6.055$                 & $6.055$                 &$3.289$		    & $3.289$                  & $3.289$                 & $3.289$\\
\hline
\end{tabular}
\label{tab:cjet-models}
\end{table*}

\subsection{Jets}
\label{sec:jets}

The delay between a successful SN explosion and the subsequent
generation of a relativistic jet from the central engine is not
completely known. Thus, we discuss next several possibilities.
\cite{Vietri_&_Stella_1998} formulated the so-called supranova model,
which predicts the formation of a supramassive NS (SMNS), i.e. a NS
stabilized by centrifugal forces at a mass exceeding the limit for
non-rotating stars \citep{Baumgarte_Shapiro_Shibata_2000}.  In the
supranova model, the timescale of energy loss and subsequently
collapse to BH for the SMNS is of the order of years, which seems to
be in conflict with the nearly simultaneous observations of GRBs
associated to SNe (see \citealt{Konigl_2004} and references
therein). However, as pointed by \cite{Guetta_&_Granot_2003}, the
SN/GRB delay time may span a wide range of values, from near
coincidence (similar to the collapsar model)\footnote{Rigidly rotating
  stars can collapse in about one orbital period
  \citep*{Shibata_Baumgarte_Shapiro_2000}.}  to even years. This broad
range of delays may reconcile the supranova model with SN/GRB almost
simultaneous detections, as well as with cases in which a SN
counterpart is not observed together with a GRB event
\citep{Woosley_&_Bloom_2006}.

Specifically for the model 35OC, the General Relativistic Hydrodynamic
core-collapse simulations of \cite{OConnor_&_Ott_2011} predict BH
formation times after core bounce in the range $[0.84,2.7]\,$s,
depending on the nuclear EoS considered. These calculations were done
including a simplified neutrino leakage scheme, unable to drive a SN
explosion. More recently, \cite{Obergaulinger_Aloy__MNRASL__2017},
including a much more elaborated neutrino transport method and
magnetic fields as indicated by stellar evolution calculations, find
that the BH formation time after bounce can be larger than the upper
bounds estimated in \cite{OConnor_&_Ott_2011}. In some models it is
even likely that a BH does not form at all. In the cases in which the
BH forms, it is necessary to wait a bit more until energy can be
efficiently extracted from the central engine, since an accretion disc
must be generated, which may take a few seconds after the BH is
formed. Furthermore, the ram pressure of the accreting matter will be
too high to allow for jet launching until the polar regions accrete so
much mass that their density decreases below $\sim 10^6\,$g\,cm$^{-3}$
\citep{MacFadyen_Woosley_ApJ_1999__Collapsar}. Altogether, the time
elapsed between core bounce and jet formation is an uncertain quantity
on the order of several seconds, as long as or even above
$\sim 5$--$10\,$s.  In this work, we consider a time delay between the
SN and the jet of $t_{\rm del} = 3$\,s.  For comparison, we note that
the SN shock of model 35OC-RO of
\cite{Obergaulinger_Aloy__MNRASL__2017} has already reached
$\simeq 3\times 10^{9}\,$cm at $1.6$\,s after the core of model 35OC
bounces, which is before BH formation.

We assume that a relativistic jet has been already formed inside our
inner boundary, which is shifted to $R_{\rm inj} (\emph{Jet}) = 2 R_0$
in the second (2D) step of our simulations (see
Sec.\,\ref{sec:model}). We note that the injection radius of the jet
and of the 1D SN ejecta differ.  We inject a conical flow through the
innermost boundary of our computational domain in the medium left
behind by the SN ejecta.  During the 1D step of our models, the SN
ejecta crosses $R_{\rm inj} (\emph{Jet})$ after
$\Delta t_{\rm SN}\simeq 0.8$\,s from its numerical injection (deeper
inside the star, at $R_{\rm inj} (\emph{SN})$). Furthermore, we let it
evolve for another period of time equal to $t_{\rm del}$ inside the
computational grid until it is sufficiently far from the jet injection
nozzle. Then we map the rest-mass density, pressure and radial
velocity profiles of the 1D SN ejecta evolution as initial conditions
for the simulations with jets (models J3) and remap them into a new 2D
spherical grid. All in all, the SN ejecta has been travelling outwards
for a time $\Delta t_{\rm SN}+t_{\rm del}\simeq 3.8\,$s before the
beginning of the 2D step of our simulations. As a calibration, we also
consider the case in which no ejecta have not been injected previously
(model J0).

The SN ejecta reduces the enclosed mass below the jet injection
boundary since, (1) it injects a mass $M_{\rm SN}$, which we assume is
extracted from the excised region below $R_{\rm inj} (\emph{SN})$, and
(2) it plows part of the stellar mass in the region between $R_0$ and
$2 R_0$. Therefore, in models J3 the inner mass enclosed below
$R_{\rm inj,j}$ is reduced to
$M_{\rm in} \text{(J3)} \simeq M_{\rm in} - M_{\rm SN}\approx 3.289
M_\odot$.

We have set isotropic equivalent energies of
$E_{\rm iso,j} = 10^{54}\,$erg (models J0a and J3a) and
$5\times10^{53}$\,erg (model J3b).  From
Fig.\,\ref{fig:energy-threshold}, we see the largest of the latter
values roughly corresponds to the threshold energy for model 35OC at
$R_{\rm inj,j}$. Imposing that the half-opening angle of the jet be
$\theta_{\rm j} = 2^\circ$, the true jet energies become
$E_{\rm j}=(1-\cos{\theta_{\rm j}})/2\times E_{\rm iso,j}\simeq
3\times 10^{50}$\,erg (J0a and J3a) and $1.5\times 10^{50}$\,erg
(J3b).  The jets are injected for $t_{\rm inj, j} = 20$\,s, so that
their isotropic luminosities are
$L_{\rm iso,j}=3.8\times 10^{52}\,$erg\,s$^{-1}$ (J0a and J3a) and
$1.9\times 10^{52}\,$erg\,s$^{-1}$ (J3b) and the true luminosities are
$L_{\rm j}\simeq 1.1\times 10^{49}$\,erg\,s$^{-1}$ (J0 and J3a) and
$5.7\times 10^{48}$\,erg\,s$^{-1}$ (J3b).  In all the models the
initial jet Lorentz factor $\Gamma_{\rm j} = 5$ and specific enthalpy
$h_{\rm j} = 20$ translate into an asymptotic Lorentz factor
$\Gamma_{\infty,0} = h_{\rm j} \Gamma_{\rm j} = 100$.  See a summary
of the parameters of each model in Tab.\,\ref{tab:cjet-models}.

We point out that \cite{Nagakura_etal_2011ApJ...731...80N} have also
studied how the earlier emergence of a shock can influence the
propagation (and emission) of a jet. In the work of
\citeauthor{Nagakura_etal_2011ApJ...731...80N} an ongoing shock arises
naturally after gravitational collapse is stalled in the star envelope
by the effect of centrifugal forces (i.e. it is not the SN shock
resulting from the core collapse, which in
\citeauthor{Nagakura_etal_2011ApJ...731...80N}'s model is assumed to
have been swallowed by the central BH). Furthermore, we note, that the
progenitor model employed in \cite{Nagakura_etal_2011ApJ...731...80N}
is not directly a progenitor resulting from a stellar-evolution
code. Instead, the authors build their own rotating equilibrium
configuration to closely mimic the density distribution of model 16TI
from \cite{Woosley_&_Heger_2006}. This model corresponds to a metal
poor star that has a ZAMS mass of $16M_\odot$, and rotates rapidly.

\section{Simulations} 
\label{sec:simulations}

In this section we show and discuss the results of the numerical
simulations employed to assess the analytic luminosity threshold found
in Sec.\,\ref{sec:threshold}. We have run two series of jet models
with different energies at the injection point: (1) the J0 series, in
which no SN ejecta was injected previously, and (2) the J3 series, in
which it was. We have checked whether jets with luminosities close to
the threshold set by Eq.\,(\ref{eq:Lj}) are able to break out the
progenitor star, at least in a time $t \lesssim t_{\rm inj, j}$.

\begin{figure}
\includegraphics[width=0.48\textwidth]{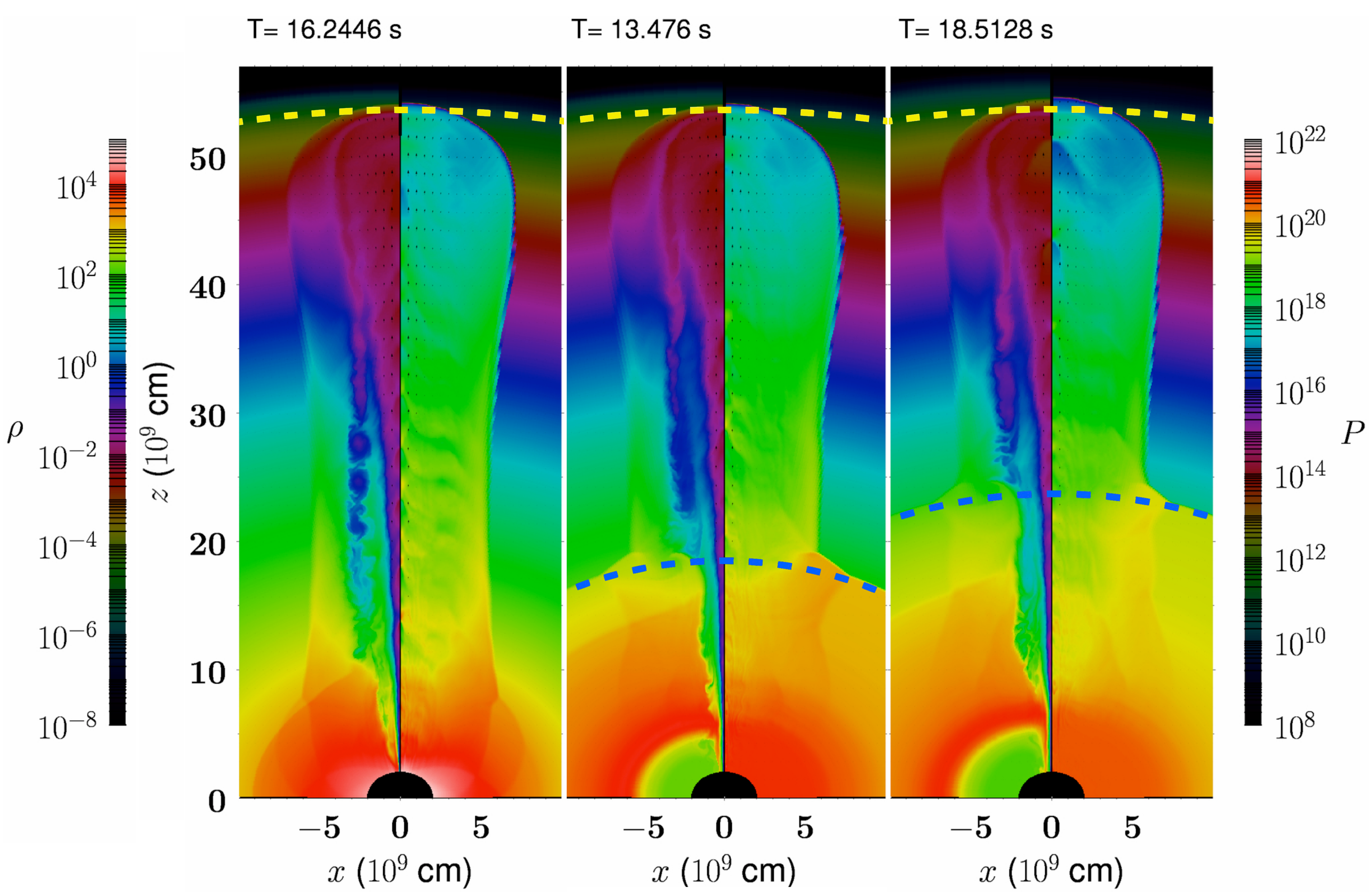}
\caption{Snapshots of rest-mass density (left side of each panel) and
  pressure (right side of each panel) in the laboratory frame of
  models J0 (left panel), J3a (central panel) and J3b (right panel) at
  breakout from the stellar surface (times are annotated above each
  panel). The rest-mass density (palette on the left side of the
  figure) and pressure (palette on the right side of the figure) are
  in CGS units. $R_\star$ and the radius, $R_{\rm FS, SN}$, reached by
  the SN forward shock (FS) are marked with yellow and blue dashed
  lines, respectively. (the time since injection is written above each
  panel)}
\label{fig:jet-models-breakout}
\end{figure}

\begin{figure}
\centering
\includegraphics[width=0.45\textwidth]{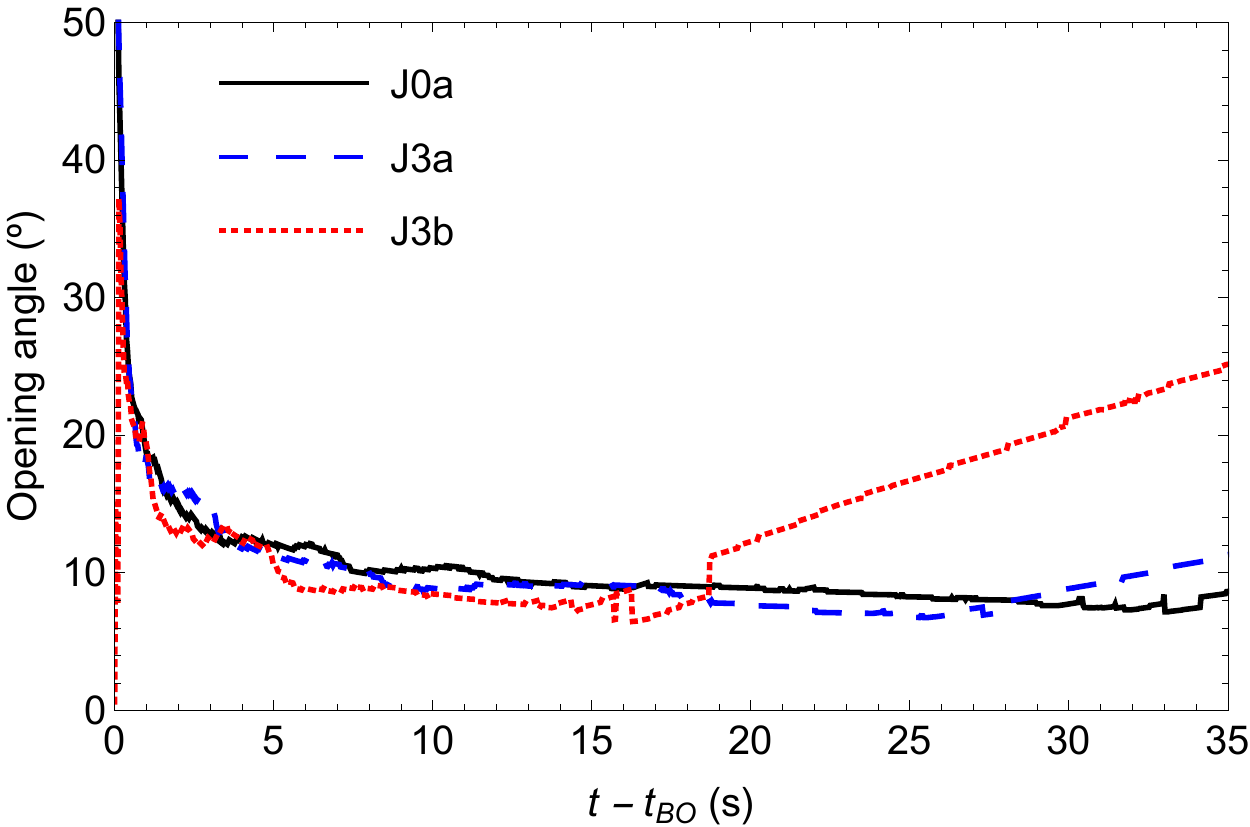}
\caption{Time evolution, in the laboratory frame, of the opening angle
  of the jet after breakout of the surface of the star for models J0a
  (black solid), J3a (blue dashed) and J3b (red dotted). The practical
  measurement of $\theta_{\rm BO}$ is rather involved around the time
  when the jets breakout and, therefore, the left part of the plot
  until $t-t_{\rm BO}\sim 2\,$s does not represent the actual jet
  half-opening angle, but instead, the jet's forward shock angular
  extension. Models J0b, J0c, J3c and J3d are not show in the figure
  for different reasons. Model J0b displays a cocoon breakout and its
  beam does not emerge out of the stellar surface, hence we cannot
  reliably measure $\theta_{\rm BO}$. Model J0c does not even enter
  successfully the grid. Model J3c displays a very quick lateral
  expansion by the time it arrives to the stellar surface and the
  values of $\theta_{\rm BO}$ cannot be measured accurately (but they
  are much larger than those of model J0a). Model J3d is chocked in
  the SN ejecta and it does not properly breakout. }
\label{fig:thetaBO}
\end{figure}

Figure\,\ref{fig:jet-models-breakout} shows those models which break
out the star in a time $t < t_{\rm inj, j}$. They correspond to the
jet models J0a and J3a and J3b. We note that model J0a possesses an
intrinsic luminosity slightly above $\Ljth$. This is also the case of
models J3a and, especially, of J3b, whose luminosity threshold is
smaller than for model J0a because of the action of the SN ejecta
(compare the blue and magenta lines in
Fig.\ref{fig:energy-threshold}). In all these models, the jet remains
well collimated from the injection until they break out of the surface
of the star. Indeed, the jets of models J3a and J3b have overtaken the
SN forward shock (FS) wave, as can be seen in the mid and right panels
of Fig.\,\ref{fig:jet-models-breakout}. After an initial transient
phase (following the jet breakout) in which $\theta_{\rm BO}$ cannot
be reliably estimated, the jet opening angle decreases and settles to
a value $\sim 9^\circ - 11^\circ$ in all jet cases
(Fig.\,\ref{fig:thetaBO}). These values are $\gtrsim 5\theta_j$, i.e.,
substantially larger than the initial opening angle of the jet. We
note that in Fig.\,\ref{fig:thetaBO}, the time range is limited to be
of the order of $t_{\rm inj}$. After that time the injection
luminosity quickly declines and the measurement of $\theta_{\rm BO}$
is polluted by the fact that the beam of the jet becomes only mildly
relativistic. The monotonical increase observed, e.g., in model J3b
$\sim 20\,$s post-breakout is, thereby, an artifact resulting from the
operative criterion employed to measure it.

\begin{figure}
\includegraphics[width=0.49\textwidth]{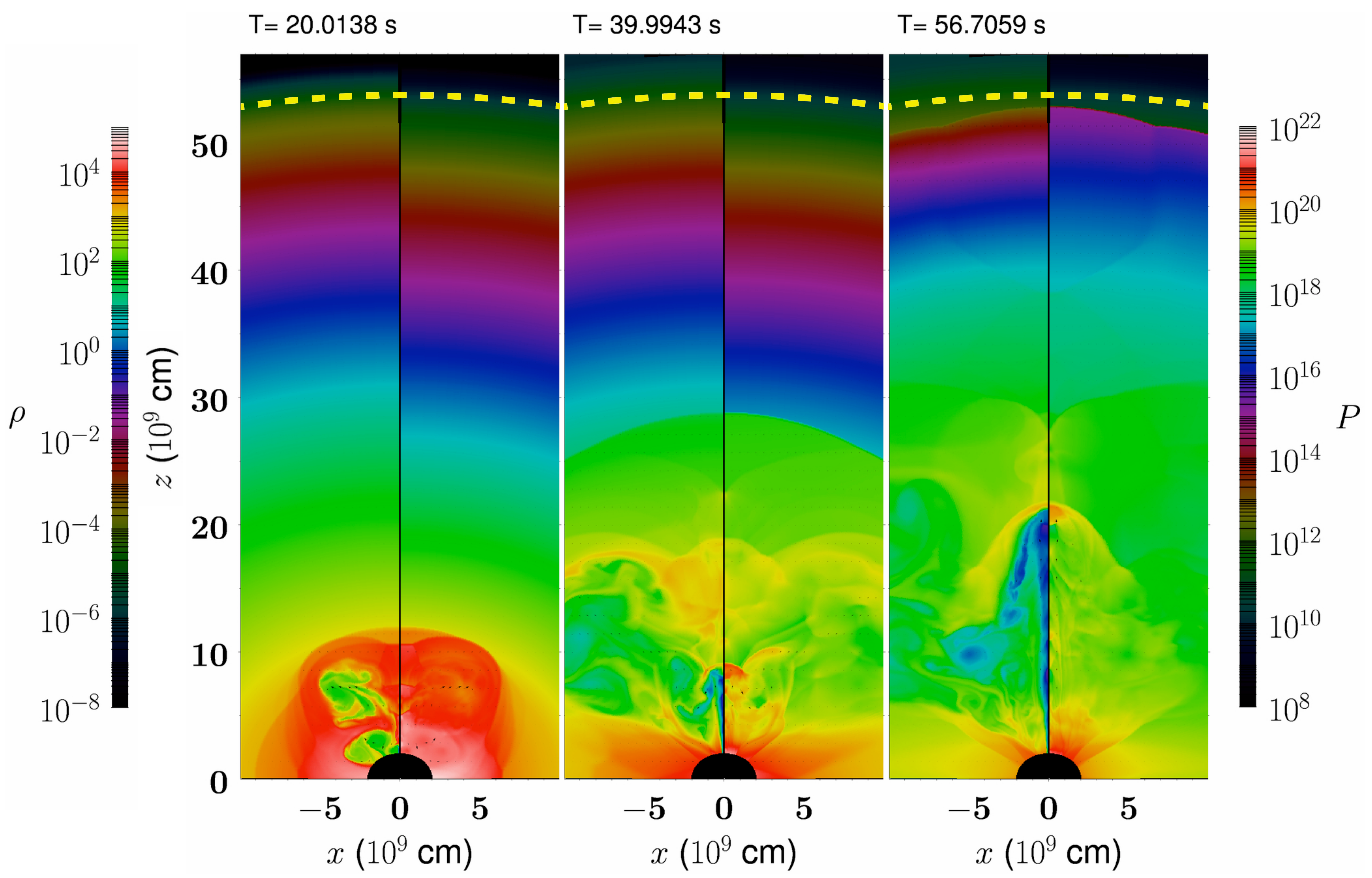}
\caption{Snapshots of the evolution of model J0b. The quantities
  displayed are the same as in
  Fig.\,\ref{fig:jet-models-breakout}. Note that depicted times are
  all longer than $t_{\rm inj,j} = 20$\,s (the time since injection is
  written above each panel). }
\label{fig:jet-models-J0}
\end{figure}

Three snapshots of the evolution of model J0b, with an injection
luminosity $\sim 2$ times below the threshold of Eq.\,(\ref{eq:Lj})
are shown in Fig.\,\ref{fig:jet-models-J0}.  In this case, the jet is
successfully injected in the computational grid, but it develops a
large, quasi-spherical, subrelativistic cocoon within the star and it
is much less collimated than the most luminous model of the J0
series. Moreover, the beam of the jet remains trapped by the thick
cocoon surrounding it. Instead of a jet breakout, we find a
\emph{cocoon breakout} from the stellar surface, which happens at a
time $t \sim 57 \text{\,s}\,> t_{\rm inj,j}$ and traps the jet
itself. We have checked that for the jet model J0c, with a luminosity
$\sim 4$ times below $\Ljth$, injection directly fails and the jet
does not even progress inside the computational grid. From the
exploration of the J0 series of models we find that model J0b is a
borderline case between successful jet injection and jet injection
failure.
\begin{figure}
\includegraphics[width=0.49\textwidth]{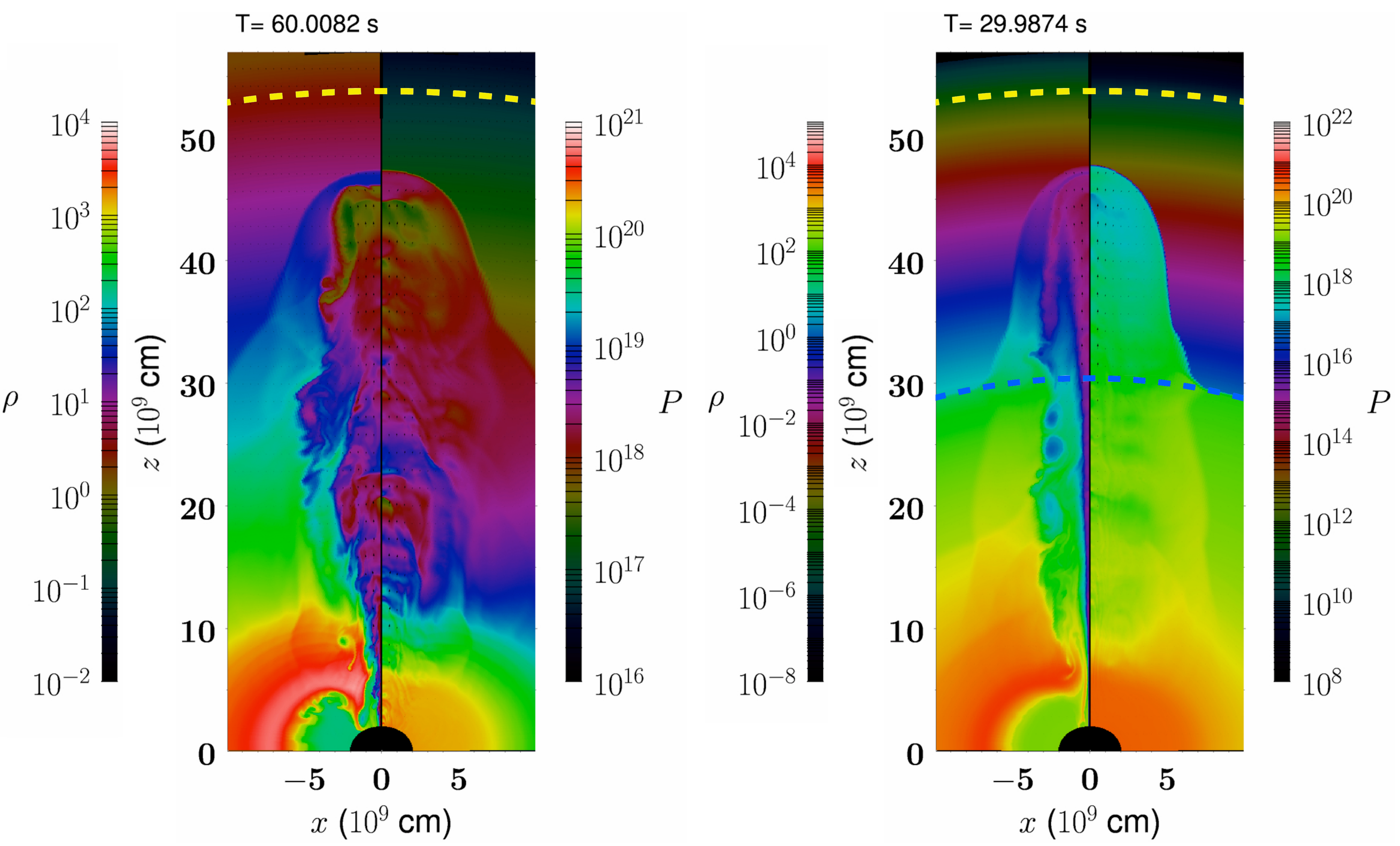}
\caption{Snapshots of the evolution of models J3d (left panel) and J3c
  (right panel). The quantities displayed are the same as in
  Fig.\,\ref{fig:jet-models-breakout}, but the color scales of the
  left panel are different for better visualization.  On the left
  panel the FS of the SN ejecta is already outside the represented
  length. Note that depicted times are different and longer than
  $t_{\rm inj,j} = 20$\,s (the time since injection is written above
  each panel).}
\label{fig:jet-models-J3}
\end{figure}

In addition to the fact that the passage of the SN ejecta may lower
the pressure and the rest-mass density in the vicinity of the jet
injection nozzle, it also drives a radial outward motion of the
stellar matter with a subrelativistic speed $v_{\rm r, a}$. If
$v_{\rm r,a} \sim v_{\rm h}$, the value of $\Ljth$ may be further
decreased, since the ram pressure on the jet's head is significantly
reduced. Thus, we have considered two additional members of the J3
series with luminosities 4 (J3c) and 10 (J3d) times than the
luminosity threshold found for the reference model J0a
(Tab.\,\ref{tab:cjet-models}), slightly below the threshold predicted
analytically for the J3 series of models
(Fig.\,\ref{fig:energy-threshold}; magenta line).  The model J3c,
hosting an equivalent isotropic energy twice smaller than
$E_{\rm iso}^{\rm thr}(\text{J3b})$ at $R_9=2$, breaks of out the star,
but after a time $t \sim 3 t_{\rm inj,j}$
(Fig.\,\ref{fig:jet-models-J3}; right panel). On the other hand, in
the model J3d, with $E_{\rm iso} = 10^{53}$\,erg (a factor of 5 less
than $E_{\rm iso}^{\rm thr}(\text{J3})$) the SN ejecta breaks out the
star before the jet does it (Fig.\,\ref{fig:jet-models-J3}; left
panel). In fact, the jet of model J3d is trapped within the SN ejecta
and fails to catch up with them because the latter propagates faster
once they enters the EM.

\section{Discussion and conclusions}
\label{sec:conclusions}
%
Aiming to explore the dynamics of relativistic jets, which may result
into events compatible with the phenomenology of llGRBs and GRB-SNe,
we have studied both analytically and numerically the conditions under
which low-luminosity jets may break out of a potential stellar
progenitor of the former type of events. In order for a jet, i.e. a
supersonic flow to form at a pre-stablished injection nozzle, the
luminosity must be above a certain threshold depending on the
conditions within the progenitor at the injection point. In order to
assess the influence of the progenitor conditions several
considerations follow.  The collapse of the inner core of low
metallicity, fast-rotating, massive stars may produce the conditions
to generate the central engine of a GRB as well as a successful SN
explosion.  Our numerical models are set up assuming that unspecified
processes (e.g. magnetohydrodynamic stresses or neutrino heating) have
generated collimated outflows inside the core of the collapsed star.
The propagation of a SN shock produces a drastic change in the
structure of the star undergoing the SN explosion. As the SN ejecta
sweep the stellar mantle, they modify the medium in which a GRB jet,
produced by the central engine, shall propagate. We have justified
that the delay between the generation of the SN ejecta and the
trailing GRB jet should probably be of, at most, a few seconds on the
basis of theoretical models and recent simulations of the formation of
the central engine of GRBs.  \cite{Obergaulinger_Aloy__MNRASL__2017}
show that a SN explosion may develop within less than a second after
core bounce.  While the shock wave propagates outwards, the conditions
at the center might lead to the launching of a GRB jet after a further
delay of the order of a few seconds.  However, since we expect the GRB
jet to be substantially faster than the SN ejecta, the former will
eventually catch up the later, drill its way trough the expanding SN
shocked layer and proceed later through the rest of the stellar
progenitor (unless the jet/SN ejecta interaction catastrophically
hinders the ulterior jet propagation).

In this paper, we have also assessed the existence of the
aforementioned theoretical luminosity threshold by means of
multi-dimensional, special relativistic, hydrodynamics simulations.
Our jets are injected at the polar axis through the inner radial
boundary situated at $R_{\rm inj,j} = 2 \times 10^{9} \, \mathrm{cm}$
with a half-opening angle of $\theta_{\rm j} = 2^\circ$. Jets with
smaller injection half-opening angles are hardly compatible with
hydrodynamic or MHD generation and collimation processes. Jets with
larger injection half-opening angles are endowed with larger
luminosities in our model set up, and thus, they are very far from the
conditions expected to generate llGRBs\footnote{For a thought
  discussion on the effects of the jet injection half-opening angle
  see, e.g. \cite{Mizuta_etal_2006ApJ...651..960M}.}. Two of our jet
models (J0a and J3b) have been chosen with luminosities only a bit
above to threshold at $R_9 = 2$ in the pre-supernova model 35OC while
others (J0b, J0c, J3c and J3d) fall below the aforementioned
luminosity threshold. We have found numerically that jets with
luminosities below the threshold of
Eq.\,(\ref{eq:luminosity-threshold}) failed in breaking out of the
star.


Beyond the theoretical point that insufficiently powerful jets fail to
even travel through the stellar mantle, the foremost relevant point is
addressing how the intrinsic jet luminosity thresholds we have
obtained translate into thresholds in the equivalent isotropic
luminosity of the jets and, hence, on the expected
$L_{\rm \gamma, iso}$. In order to translate the intrinsic jet
luminosities at the jet injection point into equivalent isotropic
quantities it is necessary, first of all, a reliable estimation of the
outflow opening angle after the jet breaks out of the surface of the
star. Two basic possibilites have been considered: either the jet
opening angle is basically the same as the initial jet opening angle
or much larger than that. The former alternative follows from the
results of \cite{Mizuta_&_Ioka_2013}. The latter from our own
results. Assuming that the outflow opening angle remains as small as
$\theta_j$, models J0a and J3a would have
$E_{\rm iso, j} = 10^{54}$\,erg ($L_{\rm iso, j} =5\times10^{52}$\,erg
s$^{-1}$), while model J3b would yield
$E_{\rm iso, j} = 5 \times 10^{53}$\,erg
($L_{\rm iso, j} =2.5\times10^{52}$\,erg s$^{-1}$).
We have tried to push down the previous values to bring them as close
as possible to the observed range in llGRBs
($L_{\rm \gamma, iso} \sim 10^{46}$--$10^{48}$\,erg), but values below
2-3 times $10^{51}\,$erg\,s$^{-1}$ are unattainable if we aim to
launch a jet that is not choked inside of the star
(Fig.\,\ref{fig:energy-threshold}) in the typically assumed
progenitors of llGRBs.  We must bear in mind that the hydrodynamic
isotropic luminosities do not directly correspond to the $\gamma$-
and/or X-ray radiation, since part of this energy may be still stored
in the form of a thermal energy reservoir that can be released on
longer time scales, over lower observational frequencies (e.g., in the
radio band) or converted to kinetic energy of the outflow.  Moreover,
part of the injected energy can be dissipated by its interaction in
the progenitor.  Following the arguments of \cite{Mizuta_&_Ioka_2013},
the energies considered here would need a tiny efficiency factor for
conversion of hydrodynamic energy into radiation,
$\epsilon_\gamma\lesssim 10^{-5}$, to lie in the llGRB regime.

Our models show that jets develop opening angles after breakout
$\theta_{\rm BO} \gg \theta_j = 2^\circ$, so that the luminosity
condition relaxes to Eq.\,(\ref{eq:luminosity-threshold2}) (in that
equation we assumed $a \sim 3$).  This conclusion is in contrast with
the findings of \cite{Mizuta_&_Ioka_2013}, but does not necessarily
contradict the results of the latter authors, since, our setup is
different from that of \citeauthor{Mizuta_&_Ioka_2013} in many
aspects. For instance, we employ model 35OC as stellar progenitor,
which is more extended and massive than model 16TI employed by the
former authors. We consider much lower luminosity jets
($L_{\rm j}\gtrsim 1.5 \times 10^{49}$\,erg\,s$^{-1}$) than
\citeauthor{Mizuta_&_Ioka_2013} (who use
$L_{\rm j}=10^{50}$\,erg\,s$^{-1}$). We inject jets with an asymptotic
Lorentz factor $\Gamma_{\infty,0}=100$, while
\cite{Mizuta_&_Ioka_2013} use a much larger value ($\simeq
538$). Another differences are the jet injection angle and time.
\cite{Mizuta_&_Ioka_2013} use $\theta_{\rm j}\simeq 4.6^\circ$, and
$t_{\rm inj,j}=12.5$\,s, while we use $\theta_{\rm j}\simeq 2^\circ$,
and $t_{\rm inj,j}=20$\,s. Finally, the inference of the former
authors on the jet opening angle is not time independent.
$\theta_{\rm BO}$ could depend on the setup of the circumburst
medium. As can be seen from their Fig.\,9, there is some trend for the
jet breakout angle to increase with time. Due to the larger values of
$\theta_{\rm BO}$ of our models, jets with the same injection
luminosities as we have set up produce lower equivalent isotropic
luminosities than in \citeauthor{Mizuta_&_Ioka_2013}. This means that,
efficiencies of $\epsilon_\gamma\sim 1\%$ could give rise to
$\gamma$-ray luminosities in the range observed for the llGRB
population in our jet models.

The duration of jet injection is a free parameter
of our simulations. It has been set to $t_{\rm inj,j} = 20$\,s,
which is more typical for standard long GRBs rather than for llGRBs,
which show durations from hundreds to thousands of seconds. However,
the existence of an injection luminosity threshold is
  independent of the duration of the energy injection. Noteworthy,
the breakout times of those models with
$L_{\rm j}\gtrsim \Ljth$ (especially the one
measured for model J3b,
$t_{\rm BO} \text{(J3b)} = 18.51$\,s) are close to the
injection time (i.e. $t_{\rm BO} \lesssim t_{\rm
  inj,j}$). \cite{Bromberg_Nakar_Piran_2011} suggested that jets with
 $t_{\rm BO}>t_{\rm inj,j}$ would fail in breaking out
of the star as they would be choked by the stellar
envelope.  We have examples of jets with
$t_{\rm BO} > t_{\rm inj,j}$ (e.g. J3c) that still can break out of its progenitor
star. However, we do not expect such jets to produce
very luminous events since the jet/SN ejecta interaction is strong and
may dissipate a large fraction of the jet internal energy.
In any case, the estimated analytical threshold for
well posed jet injection is related with the breakout
time. Precisely, we find $t_{\rm BO} > t_{\rm inj,j}$
when $L_{\rm j}$ is a
  factor $\sim 2$ smaller than $ \Ljth$. Furthermore, we find that
jets with an intrinsic luminosity smaller than $\sim \Ljth / 4$
get trapped within the stellar progenitor or within the SN ejecta, and
that jets with
  $L_{\rm j} \lesssim \Ljth/5$ are not even properly injected in the
numerical grid.

Remarkably, we have found with our numerical tests that the
  simple analytic estimate for $\Ljth$ over-predicts the existence of
  a luminosity threshold by a factor $\sim 2-3$. The analytic estimate
  is more accurate for models without a SN ejecta perturbing the
  stellar structure. This is expected, since we have neglected the
  motion of the ambient medium to estimate it, and the SN ejecta
  drives the stellar layers close to the jet injection nozzle radially
  outward, which reduces the ram pressure on the head of the jet. In
  our models without a SN ejecta the analytic threshold is still a bit
  larger than found numerically. This likely due to the fact that we
  have neglected the gravitational pull in the estimation of the jet's
  head velocity. For powerful relativistic jets, the latter assumption
  is well justified. In our case, with weaker relativistic jets, whose
  head moves only slightly supersonically with respect to the stellar
  medium, the accuracy of the forme assumption is not as good. Over
  and above all these considerations, it remains true that our
  estimated $\Ljth$ is a good proxy for the minium hydrodynamic
  luminosity attainable by relativistic jets successfully propagating
  inside likely progenitors of llGRBs.

  Finally, we point out that the existence of luminosity thresholds
  for the injection of a jet (based upon the necessity of producing a
  supersonic outflow) is not restricted to the specific context we
  have addressed here. The arguments employed to derive $\Ljth$, also
  hold qualitatively in other scenarios. Among them, we outline the
  injection of low luminosity jets in the remnant left behind the
  merger of neutron stars. In full analogy with our findings in the
  present paper, this criterion should be used in combination with the
  standard assumption that the jet injection time must be sufficiently
  long for the jet to break out of the merger ejecta
  \citep{Moharana_2017MNRAS.472L..55}. We defer for a future work
  exploring this possibility.

\section*{Acknowledgements}

We acknowledge the support from the European Research Council (grant
CAMAP-259276), and the partial support of grants AYA2015-66899-C2-1-P
and PROMETEO-II-2014-069. We thankfully acknowledge the computer
resources (\emph{Llu\'is Vives} supercomputer), technical expertise
and assistance provided by the Servei de Informatica at the University
of Valencia and the Spanish Supercomputing Network (grants
AECT--2016-1-0008, AECT-2016-2-0012, AECT-2016-3-0005,
AECT-2017-1-0013, AECT-2017-2-0006, and AECT-2017-3-0007).




\bibliographystyle{mnras}
\bibliography{bibliography} 



\appendix

\section{The `piston' model}
\label{sec:piston}

The 1D SN ejecta is injected
  from the start of the simulation until a final time of
  $t_{\rm SN}$ with a piston-like model in which we
set both the energy and the mass fluxes, $\dot E_{\rm SN}
= E_{\rm SN} / t_{\rm SN}$ and $\dot M_{\rm SN} = M_{\rm SN} / t_{\rm SN}$ respectively, across the innermost radius of our
  computational domain, $R_0$. After $t_{\rm SN}$, the energy and mass injection are not switch off
abruptly but decay very rapidly with time ($\propto
t^{-12}$). This fast injection decline is introduced for
  numerical convenience, since the gas behind the rear end of the SN ejecta is very
    rarefied. The gradient of, e.g. density between ejecta and the
    trailing matter is very steep for an instantaneous end of the
    injection, which can cause a failure of the simulation.  This
    instability can be removed by a smooth transition at the end of
    the SN injection. The steep time decrease of the ejecta injection
  conditions has been tuned to keep injecting a negligible amount of
  energy and mass in the computational grid on the time scales of
  interest (a few seconds).

  Taking units in which $c=1$, the injection model is based upon the
  main assumption that $\Theta := p / \rho \ll 1$ (where $p$ and
  $\rho$ are the fluid pressure and density). This holds as long as
  $E_{\rm SN} \ll M_{\rm SN}c^2$, condition we guarantee setting up
  suitable values for the parameters $E_{\rm SN}$ and $M_{\rm
    SN}$. Using the \emph{TM} EoS \citep{Mignone_&_McKinney_2007},
  the square of the speed of sound, $c_{\rm s}$, is
\begin{equation}
\label{eq:soundspeedTM}
c_{\rm s}^2 = \frac{\Theta}{3 h} \frac{5h - 8 \Theta}{h - \Theta} \approx \frac{5 \Theta}{3 h}\, ,
\end{equation}
and the specific enthalpy is
\begin{equation}
\label{eq:enthalpyTM}
h = \frac{5}{2} \Theta + \sqrt{ \frac{9}{4} \Theta^2 + 1} \approx \frac{5}{2} \Theta + 1\, ,
\end{equation}
where the last approximate expressions in
  Eqs.\,(\ref{eq:soundspeedTM}) and (\ref{eq:enthalpyTM}) hold for
  $\Theta\ll 1$.

From the energy and mass flux we get that
\begin{equation}
\label{eq:ksi}
\xi  := h \Gamma = 1 + \frac{\dot{E}_{\rm SN}}{\dot{M}_{\rm SN} c^2} \, .
\end{equation}

Using the definition of the Mach number,
$\mathcal{M} = v / c_{\rm s} $, with $\beta$ the
(radial) velocity of the fluid, the bulk Lorentz factor takes the form
\begin{equation} 
\label{eq:lorentzfactormach}
\Gamma = 1 / \sqrt{1 - \mathcal{M}^2 c_s^2} \, .
\end{equation}

Using Eqs.\,(\ref{eq:soundspeedTM})--(\ref{eq:lorentzfactormach}), we arrive at 
\begin{equation}
\label{eq:thetapistonmodel}
\Theta = \frac{6 (\xi^2 - 1)}{45 + \xi^2 (10 \mathcal{M}^2 - 15)} \, ,
\end{equation}
which only depends on the parameters $\dot E_{\rm SN}$,
$\dot M_{\rm SN}$ and $\mathcal{M}$ and will always
be much smaller than 1 for the typical parameters
  of our SN ejecta set up.
The Mach number is set to ensure that the shock is supersonic at injection, $\mathcal{M} = 2$.

Once $\Theta$ is computed we recover in the following order, $h$ (Eq.\,(\ref{eq:enthalpyTM})), $\Gamma = \xi / h$ (Eq.\,(\ref{eq:ksi})), $\rho$ (from the mass flux) and $p = \Theta \rho$. Note that $\xi \geq h$ must be fulfilled in order to assure that $\Gamma \geq 1$, i.e. it will be satisfied that
\begin{equation}
\Theta \leq \frac{2}{5} (\xi - 1)\, .
\end{equation}
%

\section{Gravitational potential}
\label{sec:gravity}

The equations of relativistic hydrodynamics in axial symmetry,
  employing spherical coordinates, $\textbf{x}=$($r$, $\theta$),
  natural units ($c=G=1$) and neglecting the fluid's self-gravity are
  \citep[e.g.][]{Cuesta_2017PhD}:
\begin{equation}
\label{eq:rhdequations}
\frac{\partial \textbf{U}}{\partial t} + \frac{1}{r^2} \frac{\partial
  r^2 \textbf{F}}{\partial r} + \frac{1}{r \sin \theta} \frac{\partial
\sin \theta \textbf{G}}{\partial \theta} =  \mathbf{S_{\mathbf{U}}}.
\end{equation}
The vector of conserved quantities, $\textbf{U}$, and the $r$- and
$\theta$- components of the momentum, $\textbf{F}$ and $\textbf{G}$
respectively, are defined as
\begin{align}
\label{uvector}
\textbf{U} &= \begin{pmatrix}
D \\
S^r \\
S^\theta \\
\tau    \end{pmatrix}, \\
%
\label{eq:fluxes}
\textbf{F} &=                              \begin{pmatrix}
D v^r \\
S^r v^r + p \\
S^\theta v^r  \\
(\tau + p) v^r \end{pmatrix} \:\mbox{and} \\
%
\textbf{G} &=                              \begin{pmatrix}
D v^\theta \\
S^r v^\theta \\
S^\theta v^\theta + p \\
(\tau + p) v^\theta \end{pmatrix},
\end{align}
where $D$, $\textbf{S} = (S^r, S^\theta)$ and
  $\tau$ are the relativistic mass density, the momentum density and
  the energy density, respectively, all of them measured in the
laboratory frame and defined as a function of the primitive variables:
\begin{align}
D &= \rho \Gamma, \\
\textbf{S} &= \rho h \Gamma^2 \textbf{v}, \\
\tau &= \rho h \Gamma^2 - p - D.
\end{align}
The velocity is also measured in the laboratory frame, $\textbf{v}=(v^r, v^\theta)$, and relates to the four-velocity as
\begin{equation}
u^\mu = \Gamma (1, \textbf{v}),
\end{equation}
where
\begin{equation}
\label{bulklorentzfactor}
\Gamma \equiv \frac{1}{\sqrt{1 - \textbf{v}^2}}
\end{equation}
is the fluid (bulk) Lorentz factor. We note that the components of
both momentum density and velocity are expressed in orthonormal
spherical basis.

 In the absence of physical sources, the source term,
  $\mathbf{S_{\mathbf{U}}}$, only contains all the geometrical factors
  in 2D spherical coordinates:
\begin{equation}
\label{eq:polarsources}
\mathbf{S_{\mathbf{U}}} =   \displaystyle\frac{1}{r}                          \begin{pmatrix}
0 \\
2p + S^\theta v^\theta \\*[3pt]
 \displaystyle\frac{\cos \theta}{\sin \theta} p - S^\theta v^r \\
0 \end{pmatrix}.
\end{equation}

%
%

Gravitational effects may become relevant if the deeper regions
  of the star are included in the model.  Nevertheless, in order to
  reach hydrostatic equilibrium in the star, we need to introduce
  self-gravity for balancing the pressure gradient, particularly at
  the stellar surface. This is especially true if the time over which
  we compute the models is comparable or larger than the sound
  crossing time of the stellar radius.

We have included in MRGENESIS a Newtonian
gravitational potential, $\Phi$, in order to account for
self-gravity. Although our code is relativistic, we have chosen a
Newtonian potential for simplicity, but including some relativistic
corrections as considering in the source term $\rho_{\rm eff}:=(\rho h \Gamma^2 - p)$
instead of $\rho_{\rm eff}:=\rho$ alone. We note that a similar approach has
  been followed by \cite{Nagakura_etal_2011ApJ...731...80N}. Since we
  do not account for general relativity effects, the metric remains
  unchanged and, therefore, the influence of the potential only
  appears as an additional source term in the equations of the RHD
  (\ref{eq:rhdequations}). The new source term reads
  $\textbf{S}_{\rm new} = \textbf{S}_{\textbf{U}} + \textbf{S}_{\rm pot}$, where
  $\textbf{S}_{\textbf{U}}$ is defined in Eq.\,(\ref{eq:polarsources}) and
  $\textbf{S}_{\rm pot}$ denotes the source vector due to the inclusion of
  the Newtonian potential. In 2D spherical coordinates:
\begin{equation}
\label{eq:polarsource}
\textbf{S}_{\rm pot} = \begin{pmatrix}
0 \\
\\
- (\rho h \Gamma^2 - p) \dsfrac{\partial \Phi}{\partial r} \\
\\
- \dsfrac{1}{r} (\rho h \Gamma^2 + S^r v^r) \dsfrac{\partial \Phi}{\partial \theta} \\ 
\\
- \rho h \Gamma^2 \left( v^r \dsfrac{\partial \Phi}{\partial r} + \dsfrac{v^\theta}{r} \dsfrac{\partial \Phi}{\partial \theta} \right) \end{pmatrix},
\end{equation}
 %
Once the potential source term is
volume-averaged, we get
\begin{equation}
\label{eq:averagedpotential}
 \tilde{\textbf{S}}_{\rm pot} := \frac{1}{\Delta V} \int \limits_V
 \textbf{S}_{\rm pot} \text{ d}V = \begin{pmatrix} 
 0 \\
 \\
- (\rho_{i,j} h_{i,j} \Gamma_{i,j}^2 - p_{i,j}) \, S^r_{0,\rm pot}  \\ 
\\
- (\rho_{i,j} h_{i,j} \Gamma_{i,j}^2 + S_{i,j}^r v_{i,j}^r)
\, S^\theta_{0,\rm pot} \\
\\
- \rho_{i,j} h_{i,j} \Gamma_{i,j}^2 \, ( v_{i,j}^r S^r_{0,\rm pot} +
v_{i,j}^\theta S^\theta_{0,\rm pot} ) \end{pmatrix} ,
\end{equation}
being
\begin{equation}
\begin{split}
\label{eq:averagedpotential}
S^r_{0,\rm pot} =&  \frac{3}{\Delta r^3} \Big[ \Phi_{i+1/2, j} r_{i+1/2}^2 - \Phi_{i-1/2, j} r_{i-1/2}^2  - \Phi_{i, j}  \Delta r^2 \Big] \\
S^\theta_{0,\rm pot} =&  - \frac{3}{2} \frac{\Delta r^2 }{\Delta r^3
  \Delta \cos{\theta}} \times \\ &\Big[ \Phi_{i, j+1/2}
\sin{\theta_{j+1/2}} - \Phi_{i, j-1/2} \sin{\theta_{j-1/2}} - \\ & \: \Phi_{i, j}  \Delta \sin{\theta} \Big] ,
\end{split}
\end{equation}
where in a given cell $(i,j)$ the potential has to be known also at
the cell boundaries $i+1/2$, $i-1/2$, $j+1/2$ and $j-1/2$. In the
previous expression we have introduced the notation,
$\Delta r^2:=r_{i+1/2}^2-r_{i-1/2}^2$,
$\Delta r^3:=r_{i+1/2}^3-r_{i-1/2}^3$,
$\Delta \cos{\theta} := \cos{\theta_{j+1/2}}- \cos{\theta_{j-1/2}}$
and
$\Delta \sin{\theta} := \sin{\theta_{j+1/2}}-
\sin{\theta_{j-1/2}}$.

For uniformly spaced grids we make the simple
assumption that
\begin{align}
\label{eq:boundarypotential}
\Phi_{i+1/2, j} &= \frac{1}{2} (\Phi_{i+1, j} + \Phi_{i, j}) \\
\Phi_{i+1/2, j} &= \frac{1}{2} (\Phi_{i, j} + \Phi_{i-1, j}) .
\end{align}
and the same is done for the interface values in the $j$-direction.

The Poisson equation,
\begin{equation}
\Delta \Phi = 4 \pi \rho_{\rm eff}
 \end{equation}
defines the behaviour of the potential, $\Phi$, and its dependence on
the mass distribution. To find the solution of this elliptic equation, 
we have used the method devised in \cite{Adsuara_2016JCoPh.321..369,
  Adsuara_2017JCoPh.332..446}.  For numerical convenience, we do not
solve directly for the potential $\Phi = \Phi(r, \theta)$ but a
modified potential
$\Phi'(r, \theta) = \Phi (r, \theta) + M_{\rm in} / r$, where
$M_{\rm in}$ is the excised mass below $R_0$.  Neumann conditions for
the potential are imposed at $R_0$,
$\frac{\partial \Phi'}{\partial r} |_{R_0} = 0$, and Dirichlet
conditions at the outer end of the radial grid,
$\Phi' (R_f, \theta) = - M_T/R_f$. The total mass within the grid,
$M_T$, excludes the excised mass $M_{\rm in}$. Once $\Phi'$ is
calculated, we only have to subtract in the radial direction the
quantity $M_{\rm in} / r$ to recover the real potential, $\Phi$.

 The potential is recalculated after a number of iterations
  equal to multiples of the number of cells in the radial direction,
  $n_r$. The reason is that $n_r \Delta t$ corresponds to the light
  crossing time\footnote{In the ideal case with a
    Courant-Friedrichs-Lewy condition of CFL = 1.} in the grid and the
  potential is updated before any perturbation can cross the whole
  numerical grid.
  Between two consecutive calculations it is very likely that the
  inner mass has changed due to a non-zero mass flux across $R_0$
  (interface $1/2$). This flux can supply a non-negligible amount of
  mass to the enclosed mass $M_{\rm in}$, making it necessary to
  consider it in order to properly recompute the potential. Whether or
  not such a contribution is included can strongly influence the
  dynamics, specially in those regions close to $R_{\rm inj,j}$.  The total
  incoming mass, calculated as
  $\Delta M_{\rm in} = - \sum_j F_{1/2,j} S_{1/2,j}$, is evaluated in
  every time step in the same manner as for the conserved variables,
  i.e. it is updated in each of the Runge-Kutta steps of our time
  integration method.  In the latter formula $F_{1/2,j}$ is the mass
  flux per unit surface between cells $(0,j)$ and $(1,j)$
  and $S_{1/2,j} = 4 \pi (-\Delta \cos \theta) R_{\rm inj,j}^2$ is the
  surface of the respective boundary located at $R_{\rm inj,j}$.
  After the end of the time loop $\Delta M_{\rm in}$ is removed from
  the grid and incorporated to the inner mass.



\bsp	
\label{lastpage}
\end{document}